\documentclass{aa} 
\usepackage[dvipsnames]{xcolor} 
\usepackage{graphicx, subfigure, amsmath}
\usepackage[breaklinks=true]{hyperref} 
\hypersetup{colorlinks=true,citecolor=Blue}
\newcommand{\upperRomannumeral}[1]{\uppercase\expandafter{\romannumeral#1}}

\begin{document}

\title{Can dead  zones create structures like a transition disk?}

\author{
Paola Pinilla\inst{1}, Mario Flock\inst{2}, Maria de Juan Ovelar\inst{3}, and Til Birnstiel\inst{4}}
\institute{
Leiden Observatory, Leiden University, P.O. Box 9513, 2300RA Leiden, The Netherlands\\
\email{pinilla@strw.leidenuniv.nl}
\and Jet Propulsion Laboratory, California Institute of Technology, Pasadena, California 91109, USA
\and Astrophysics Research Institute, Liverpool John Moores University,146 Brownlow Hill, Liverpool L3 5RF, UK
\and Max-Planck-Institut f{\"u}r Astronomie, K{\"o}nigstuhl 17, D-69117, Heidelberg, Germany}

 \abstract
   {Regions of low ionisation where the activity of the magneto-rotational instability is suppressed, the so-called dead zones, have been suggested to explain gaps and asymmetries of transition disks. Dead zones are therefore a potential cause for the observational signatures of transition disks without requiring the presence of embedded planets.}
   {We investigate the gas and dust evolution simultaneously assuming simplified prescriptions for a dead zone and a magnetohydrodynamic (MHD) wind acting on the disk. We explore whether the resulting gas and dust distribution can create signatures similar to those observed in transition disks.}
   {We imposed a dead zone and/or an MHD wind in the radial evolution of gas and dust in protoplanetary disks. For the dust evolution, we included the transport, growth, and fragmentation of dust particles. To compare with observations, we produced synthetic images in scattered optical light and in thermal emission at mm wavelengths.}
   {In all models with a dead zone, a bump in  the gas surface density is produced that is able to efficiently trap large particles ($\gtrsim 1$~mm) at the outer edge of the dead zone. The gas bump reaches an amplitude of a factor of $\sim5$, which can be enhanced by the presence of an MHD wind that removes mass from the inner disk. While our 1D simulations suggest that such a structure can be present only for $\sim$1\,Myr, the structure may be maintained for a longer time when more realistic 2D/3D simulations are performed. In the synthetic images, gap-like low-emission regions are seen at scattered light and in thermal emission at mm wavelengths, as previously predicted in the case of planet-disk interaction.}
   {Main signatures of transition disks can be reproduced by assuming a dead zone in the disk, such as gap-like structure in scattered light and millimetre continuum emission,  and a lower gas surface density within the dead zone. Previous studies showed that the Rossby wave instability can also develop at the edge of such dead zones, forming vortices and also creating asymmetries.}


\titlerunning{Can dead  zones create  transition disk like structures?}
\authorrunning{P.~Pinilla et al.}
\maketitle

\section{Introduction}     \label{introduction}

Transition disks have dust-depleted inner regions that were first identified by the weak near- and mid-infrared excess of their spectral energy distributions \citep[SEDs,][]{strom1989, skrutskie1990}. In the past decades, some of these dust gaps/cavities have been resolved by interferometric observations at (sub-)millimetre wavelengths, showing a large diversity of morphologies,  such as rings and large-scale asymmetries \citep[e.g.][]{andrews2011, isella2013, casassus2013, marel2013, perez2014}.  In addition, other fascinating structures have been observed in transition disk at optical and near-infrared wavelengths, such as spiral arms \citep[e.g.][]{muto2012, benisty2015}, dips and/or shadows \citep[e.g.][]{mayama2012, avenhaus2014}. The combination of both scattered-light and mm emission has provided several examples that demonstrate that small and large particles have different radial distributions \citep[e.g.][]{garufi2013, pinilla2015}. In addition, gas inside the dust gaps of these disks has been detected, and this is in some cases also depleted but by a smaller factor than the millimetre dust \citep[e.g.][]{carmona2014, perez_s2015}. The drop in the gas surface density inside the mm cavities can be as large as a factor of 1000 \citep{marel2016}. Another interesting observational aspect of transition disks is that the hosting stars seem to be depleted in refractory elements compared to full classical disks, suggesting that these elements have been locked (or trapped)  in the outer part of the disks \citep{kama2015}. 

Embedded massive planets ($\gtrsim0.1-1~M_{\rm{Jup}}$)  are a common and attractive solution to explain the diverse morphologies of transition disks at different wavelengths  \citep[e.g.][]{zhu2012, pinilla2012, pinilla2015a, ataiee2013, dejuanovelar2013, dong2015, pohl2015}. Nonetheless, these models may contradict current exoplanet statistics. \cite{brandt2014} presented a uniform Bayesian analysis of a large sample of possible companions around nearby stars (250 targets at a distance of 5-130~pc), covering a wide range of spectral type (from late-B to mid-M stars) and ages (10-250~Myr), and concluded that  only 1-3\% of stars host massive planets ($5-70~M_{\rm{Jup}}$) between 10 and 100~au \citep[similar limits are found in other studies with higher upper limits for more massive stars, see e.g.][]{vigan2012, nielsen2013, chauvin2015}. These statistics seem to be low compared to the number of transition disks in the protoplanetary disk population \citep[$\sim10-20\%$ e.g.][]{andrews2011, espaillat2014}.  As an alternative explanation, in this work we explore the possibility that dead zones  may be a potential explanation of the origin of transition disks, by modelling the simultaneous evolution of gas and dust over timescales of several million years. In addition, we also include the effect of a magnetohydrodynamic (MHD) wind that removes mass from the inner disk. Previous studies of dust and gas distributions with dead zones either consider a constant dust-to-gas ratio or the evolution of a single particle size \citep[e.g.][]{morishima2012, regaly2012, flock2015}, whereas for the dust evolution we include its transport, growth, and fragmentation. Our prescriptions for both dead zones and MHD winds are a simplified formulation of what is expected from fully magneto-rotational instability (MRI) simulations.

To sustain accretion onto the central star, angular momentum must be transported outward in protoplanetary disks \citep[see][for a review]{armitage2011}. Different sources of angular momentum transport have been suggested to play a role for the disk evolution, including MHD winds \citep[e.g.][]{blandford1982, suzuki2009, bai2016}, self-gravity \citep[e.g.][]{lin1987, lodato2004, vorobyov2009}, and hydrodynamical instabilities such as baroclinic instabilities \citep[e.g.][]{klahr2003, raettig2013} and vertical shear instability \citep[e.g.][]{urpin1998, nelson2013}. One preferred mechanism for angular momentum transfer is the MHD turbulence from MRI, which originates from the magnetic tension between adjoining fluid elements of the disk \citep[e.g.][]{balbus1991, balbus1998}. Non-ideal MHD effects, such as Ohmic resistivity, ambipolar diffusion, and Hall drift have an important impact on the coupling of the magnetic field and so on the disk dynamics \citep[e.g.][]{turner2014}. In particular, they have a significant effect at the location where the disk gas is expected to be decoupled from the magnetic field, in the well-known dead zones. 

The size and shape of dead zones depend on the ionisation degree of the gas, which can originate from thermal ionisation in the very inner parts of the disk where the temperatures are $\gtrsim 1000~$K \citep{Umebayashi1988, desch2015}, or from external ionisation from stellar X-rays, far-ultraviolet (FUV) photons, or cosmic rays from the interstellar space \citep[e.g.][]{dolginov1994, glassgold1997, sano2000}. In disk regions with a high ionisation degree, the MRI can generate sufficiently high turbulence to explain the angular moment transport and so the disk evolution \citep{fromang2006, davis2010, flock2012}. Recent local and global MHD simulations including ambipolar diffusion \citep{bai2013, gressel2015} have shown that MRI is suppressed in the bulk regions between $\sim1$~au to $\sim$10~au and that the angular momentum is mainly removed by an MHD wind. MHD winds can disperse the gas from the inside out even when a dead zone exists farther out  \citep{suzuki2009, fromang2013}. 

On the other hand, the ionisation degree of the disk (hence the shape of the dead zone) is strongly influenced by the dust surface area, and  small and/or large fractal  grains with large surface area can sweep up available ions and electrons more efficiently than large compact grains \citep{okuzumi2009}. \cite{dzyurkevich2013} investigated how fluffy aggregates can affect the structure of dead zones, finding a smooth transition between dead and active regions, and a  large diversity of sizes and shapes for the dead zone. In general, how much and where the MRI is suppressed in protoplanetary disks is a complex problem that significantly depends on different parameters such as dust properties (e.g. size, volume density, and charge) and its abundance, the magnetic field \citep[e.g.][]{simon2015}, the disk chemistry \citep[e.g.][]{perez_becker2011}, and the sources of ionisation \citep[e.g.][]{cleeves2013}. Overall, when MRI is suppressed, the rate of gas flow decreases \citep[e.g.][]{blaes1994} and therefore gas accumulates in the transition from high to low ionisation regions, and a local pressure maximum can form. This pressure bump is capable of trapping particles \citep[e.g.][]{varniere2006, kretke2007,brauer2008, dzyurkevich2010, joanna2013, ruge2016} and stops the rapid inward migration of the larger pebbles \citep[e.g.][]{weidenschilling1977}.  

For the gas and dust evolution models in this work, we use simple and independent parameterisations to include a dead zone and/or an MHD wind acting on the disk. Under these assumptions, we aim to study the potential trapping of particles and compare the results with current observations of transition disks. The paper is organised as follows. In Sect.~\ref{method} we introduce the method for studying the gas and dust evolution, the analytical prescriptions to consider a dead zone and an MHD wind, our initial conditions and assumptions, and the radiative transfer modelling. In Sect.~\ref{results} we present the main results when dead zones with different morphologies are considered and when an MHD wind is also assumed. In addition, in this section we show the synthetic images after the radiative transfer modelling. In Sect.~\ref{discussion} we discuss our results and compare them with other models for the origin of transition disks signatures, such as planet disk interactions. Finally, the summary and main conclusion of our work are laid out in Sects.~\ref{summary} and ~\ref{conclusion}, respectively.

\section{Method}     \label{method}

\subsection{Gas and dust evolution} \label{method_a}

For the gas evolution, we solve the diffusion equation obtained from the continuity equation and from the conservation of angular momentum of the gas in protoplanetary disks \citep{lust1952, lynden1974}, 

\begin{equation}
\frac{\partial \Sigma_g}{\partial t}=\frac{3}{r}\frac{\partial}{\partial r}\left[ r^{1/2} \frac{\partial}{\partial r} (\nu \Sigma_g r^{1/2})\right], 
\label{eq_gas_evo}
\end{equation}

\noindent where $\Sigma_g$ is the gas surface density and $\nu$ is the disk kinematic viscosity responsible for the angular momentum transport in the disk. Commonly, the models from \cite{shakura1973} are used for a simplified parametrisation of the kinematic viscosity, which is assumed to be 

\begin{equation}
\nu=\alpha c_s  h \quad \textrm{with} \quad h=\frac{c_s}{\Omega},
\label{viscosity}
\end{equation}

\noindent where $c_s$ is the isothermal sound speed, $\Omega$ the Keplerian frequency, $\alpha$ a dimensionless parameter ($\alpha \leq 1$), and $h$ the pressure scale height. In the models assuming a high ionisation state, we consider  $\alpha$ to be constant  and equal to $10^{-2}$, to compare with previous studies that used this value. To include a dead zone for the disk evolution, we assume that $\alpha$ radially changes as

\begin{eqnarray}
\alpha\left(\Sigma_g(r, t)\right)&=&\left(1-\tanh\left[(\Sigma_g(r, t)-\chi)/(\rm{1g~cm}^{-2})\times \varepsilon\right] \right) \nonumber \\
&&\frac{\alpha_{\rm{active}}}{2}+ \alpha_{\rm{dead}},
\label{dead_zone}
\end{eqnarray}

\noindent where $\chi$ and $\varepsilon$ control where and how steep the transition between $\alpha_{\rm{dead}}$ and $\alpha_{\rm{active}}$ is with respect to the local value of the gas surface density (see Tables~\ref{parameters}~and~\ref{assumed_models}). \cite{regaly2012} and \cite{miranda2016} used similar analytical expressions for the dead zone, but with Eq.~\ref{dead_zone},  we also implement a dependence on surface density, so that a change in surface density can switch the disk from active to dead and back.  In our models, we fit $\alpha_{\rm{active}}=2\times10^{-2}$ and $\alpha_{\rm{dead}}=10^{-4}$ to values that approximate to more realistic profiles using constraints from local and global MRI simulations: the chosen value of $\alpha_{\rm{active}}$, which is slightly above $10^{-2}$, is justified by local and global MRI simulations in well-ionised media \citep[e.g.][]{fromang2006, davis2010, simon2011, flock2012}. The value of $\alpha_{\rm{dead}}$ is still under intense research while a value of $10^{-4}$, which is representative for the turbulent component, could be seen as a lower limit, arising from hydrodynamical instabilities \citep{nelson2013, lyra2014, klahr2014}, for instance.

To add the influence of an MHD wind in the disk evolution, a loss term ($\dot{\Sigma}_{\rm{wind}}$) is included in the viscous evolution of the gas surface density (Eq.~\ref{eq_gas_evo}), which is assumed to be proportional to the local Keplerian frequency and the local gas surface density \citep{suzuki2010, ogihara2015}, such that

\begin{equation}
\dot{\Sigma}_{\rm{wind}}= -C_w \frac{\Sigma_g}{\sqrt{2\pi}}\Omega.
\label{wind_term}
\end{equation}

$C_w$  is assumed to be a dimensionless constant that represents the strength of the disk wind. This is a valid approximation for modelling an MHD wind when the net vertical magnetic field is weak \citep{ogihara2015}. The mass-loss rates from MHD winds are not well known, but are expected to be efficient mostly in the inner disk \citep[$\lesssim2-5$~au, e.g.][]{armitage2013, bai2013}, and to have values of $\sim10^{-8}\,M_\odot~$year$^{-1}$ \citep[e.g.][]{simon2013, bai2016b}. We note that we implement these two MHD effects as parametrised and independent of each other,  and they are simplifications of what is expected from MRI simulations. In particular, our implementation of an MHD wind only contributes to a mass-loss rate.  In reality, an MHD wind can carry angular momentum as well and contribute to the accretion flow \citep{armitage2013, bai2013, fromang2013}.  Our models study the effects of mass-loss and angular momentum transport independently, which is useful as a simplification, but may not be fully realistic.  

\begin{table}
\caption{Fixed model parameters}
\centering   
\tabcolsep=0.08cm                      
\begin{tabular}{c|c|c}       
\hline
\hline
Parameter &Symbol /units&  Value \\
\hline
Stellar mass&$M_{\star}[M_{\odot}]$& $1$\\
Stellar radius&$R_{\star}[R_{\odot}]$& $2.5$\\
Effective stellar temperature &$T_{\star}[K]$& $4300$\\
Initial gas density at 100 au&$\Sigma_0[ \rm{g~cm}^{-2}]$& $6$\\
Inner disk radius&$r_{\rm in}[\rm{au}]$ & $1.0$\\
Outer disk radius&$r_{\rm out}[\rm{au}]$& $200$\\
Radial grid resolution&$n_r$& $300$\\
$\alpha$ active region&$\alpha_{\rm{active}}$& $2\times10^{-2}$\\
$\alpha$ dead region&$\alpha_{\rm{dead}}$& $10^{-4}$\\
$\Sigma$ threshold of the dead zone edge &$\chi [\rm{g~cm}^{-2}]$& $[7.5, 15]$\\
Smoothness of dead-zone edge &$\varepsilon$& $[0.2, 0.4]$\\
Wind strength&$C_w$& $2\times10^{-5}$\\
Minimum grain size&$a_{\rm{min}}[\rm{cm}]$ &$10^{-4}$\\
Maximum grain size&$a_{\rm{max}}[\rm{cm}]$ &$2\times10^{2}$\\
Grain size grid resolution&$n_g$& $180$\\
Fragmentation velocity&$v_f[\rm{m~s}^{-1}]$ &10\\ 
Volume density of dust&$\rho_s[ \rm{g~cm}^{-3}]$&1.2\\
Distance to the disk&$d[\rm{pc}]$&140\\
\hline
\hline
\end{tabular}    
\label{parameters}
\end{table}

For the inner boundary condition, we considered two different tests. One is that the inner and outer value of the gas surface density is set to a floor value (taken to be $10^{-100}$). Another test is that zero-gradient condition is used for the inner boundary while for the outer boundary the floor-value condition is kept. Comparing the results of the gas distribution for the two types of boundary conditions, the main results do not show significant differences. The condition used for the results that are shown in the paper is the floor-value condition for the inner and the outer boundary.

For the dust evolution, we simultaneously modelled the transport of the dust and its growth as implemented by \cite{birnstiel2010}. We solved the advection-diffusion equation of the dust surface density ($\Sigma_d$) for each grain size, given by

\begin{equation}
	\frac{\partial \Sigma_d}{\partial t} + \frac{1}{r}\frac{\partial}{\partial r}\left( r \Sigma_d v_{\mathrm{r,d}}\right)-\frac{1}{r}\frac{\partial}{\partial r} \left(r \Sigma_g D_d \frac{\partial }{\partial r}\left[\frac{\Sigma_d}{\Sigma_g}\right]\right)=0,
  \label{eq:dustevo}
\end{equation}

\noindent where $D_d$ is the dust diffusivity, which is assumed to depend on the kinematic viscosity of the gas \citep{youdin2007}

\begin{equation}
	D_d=\frac{\nu}{1+\mathrm{St}^2}, 
  	\label{eq:diffusion}
\end{equation}

\noindent where $\nu$ is calculated at each radius as Eq.~\ref{viscosity}, which depends on $\alpha (r, t)$ (Eq.~\ref{dead_zone}). The value of $\alpha (r, t)$ is determined following the surface density evolution of gas (Eq.~\ref{eq_gas_evo}). The radial dust velocity ($v_{\mathrm{r,d}}$) is given by

\begin{equation}
	v_{\mathrm{r,d}}=\frac{v_{\mathrm{r,g}}}{1+\textrm{St}^2}+\frac{1}{\textrm{St}^{-1}+\textrm{St}} \frac{\partial_r P}{\rho_g \Omega},
\label{eq:dustvel} 
\end{equation}

\noindent where $P$ and $\rho_g$ are the gas pressure and the total gas density at the mid-plane, respectively, that is, $\rho=\Sigma_g/\sqrt{2\pi}h$ and $P=c_s^2\rho$. The dust diffusivity and the radial dust velocity depend on the Stokes number ($\textrm{St}$), which is the stopping time of the particle within the gas. In the Epstein regime ($\lambda_{\mathrm{mfp}}/a~\geq~4/9$, with $\lambda_{\mathrm{mfp}} $ being the mean free path of the gas molecules and $a$ the size of the grain size), $\textrm{St}$ is defined at the disk midplane as

\begin{equation}
	\textrm{St}=\frac{a\rho_s}{\Sigma_g}\frac{\pi}{2},
 	\label{eq:stokes}
\end{equation}

\noindent with $\rho_s$  being the volume density of a dust grain. 

The grain growth is calculated by solving the Smoluchowski coagulation equation \citep{smoluchowski1916} and assuming that particles fragment when they reach a threshold  for their relative velocities, that is, the fragmentation velocity ($v_f$), whose values depend on the dust properties and can vary from 10-80~$\mathrm{m~s}^{-1}$ for ices \citep[e.g.][]{wada2013}. For more details about the modelling of the dust evolution, we refer to \cite{birnstiel2010}.

\begin{figure}
 \centering
      	\includegraphics[width=9cm]{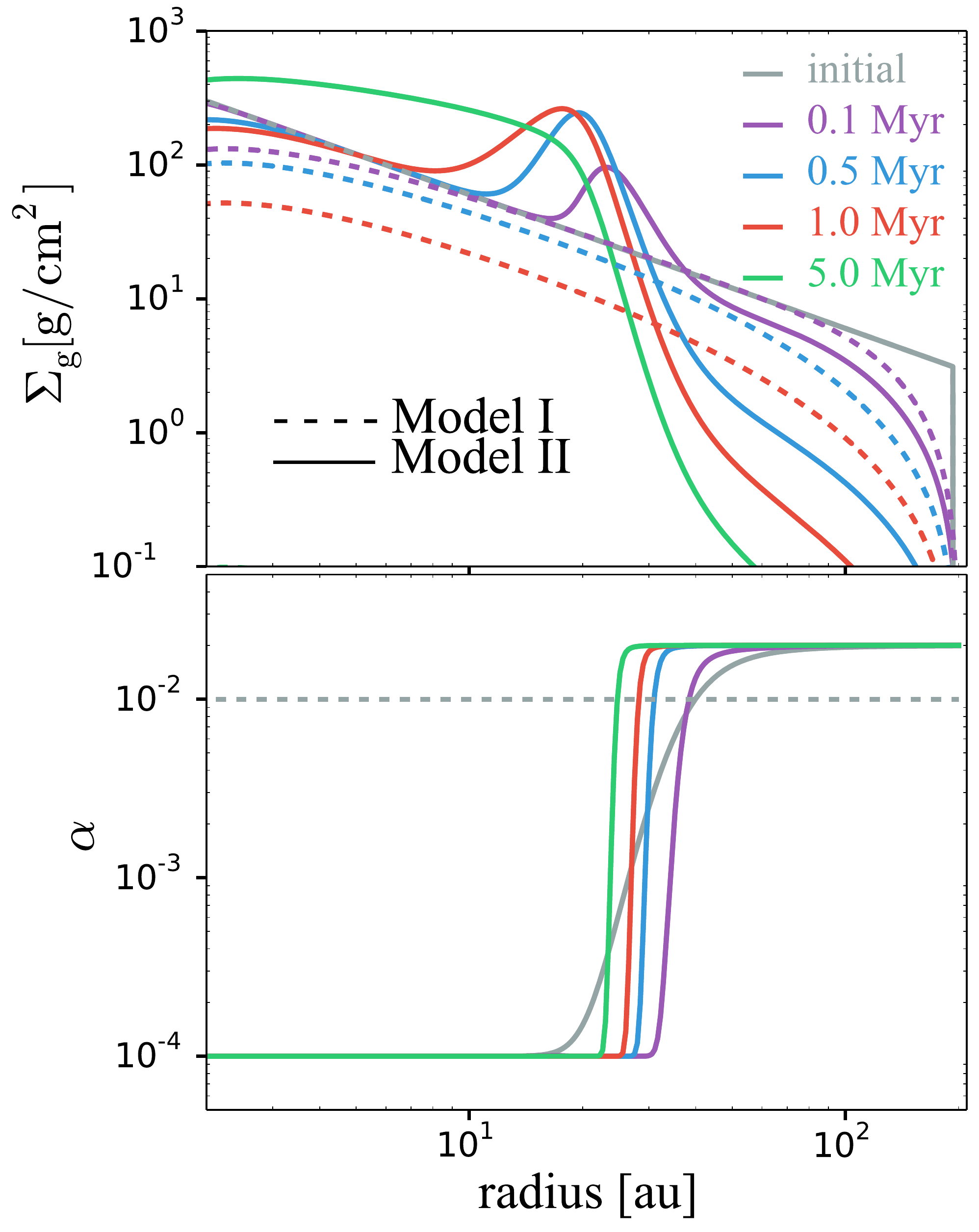}
    \caption{Evolution of the gas surface density (top panel) and $\alpha(r,t)$ profile (bottom panel) when a dead zone is assumed in the disk  (Eq.~\ref{dead_zone}, model~\upperRomannumeral{2}). For comparison, the dashed lines represent the evolution of a model without a dead zone (model~\upperRomannumeral{1}).}
   \label{model1_2}
\end{figure}

\subsection{Radiative transfer and synthetic images} \label{method_b}

To create synthetic intensity images from our models, we used MCMax which is a 2D Monte Carlo radiative transfer code \citep{min2009}. We assumed a wavelength grid from 0.1$\mu m$ to 3\,mm with 100 cells to calculate the dust temperature. We used the dust density distribution for different grain sizes previously obtained from the dust evolution models at a certain time of evolution. The settling and vertical turbulent mixing were calculated self-consistently with the values of $\alpha$ (Eq.~\ref{dead_zone}) at each radial position. We considered a mixture of  silicates,  iron sulphide, and carbonaceous  grains  for the dust composition as in  \cite{dejuanovelar2013}. We selected two wavelengths to generate the synthetic images, one optical at R band ($0.65~\mu$m) and one at thermal millimetre  emission (850$\mu$m). We calculated the Stokes Q and U images and together with the intensity maps at R band, we obtained the polarised intensity images (PI) by considering PI=$\sqrt{Q^2+U^2}$.

\subsection{Initial conditions and assumed parameters } \label{method_c}
We assumed that the initial gas surface density profile is a power law, similar to \cite{flock2015}, such that 

\begin{equation}
	\Sigma_{g, 0}=\Sigma_0 \left(\frac{r}{100~\rm{au}}\right)^{-1},
 	\label{eq:initial_gas}
\end{equation}
 
\noindent and the initial disk mass is $0.08~M_{\odot}$. As we show below, the presence of a dead zone significantly modifies the surface density profile and it will eventually deviate from this initial assumption.  If the dead zone is present throughout the early evolution of the disk, a power law disk profile as it results from viscous accretion disk models with radially constant $\alpha$ would then not be a realistic initial condition.  However, such a power law is a realistic assumption if the disk arrived at its current state from a period of much stronger accretion, either during a Class I phase, or possibly also during an FU~Ori-like strong accretion event during which the entire disk is expected to warm up and become viscously coupled over large distances \citep[e.g.][]{hartmann1998}.  If the disk then settles to the low accretion rates observed of Class II objects and in particular transitional disks,  we believe that a power law surface density is a good assumption for the initial disk state. 

\begin{table*}
\caption{Assumed models}
\centering                  
\begin{tabular}{c|c|c|c|c|c|c|c|c} 
\hline
\hline
&&&&&&&&\\      
\textbf{Model} &\upperRomannumeral{1} &\upperRomannumeral{2}&\upperRomannumeral{3}&\upperRomannumeral{4}&\upperRomannumeral{5}&\upperRomannumeral{6}&\upperRomannumeral{7}&\upperRomannumeral{8}\\
&&&&&&&&\\
\hline
Dead zone &no ($\alpha=\rm{cst}=10^{-2}$) &yes&no&yes&yes&yes (steeper)&yes (farther out)&yes\\
$\chi$&--& $15$& --& $15$& $15$& $15$& $7.5$& $15$\\
$\varepsilon$&--& $0.2$& --& $0.2$& $0.2$& $0.4$& $0.4$& $0.2$\\
\hline
MHD wind &no&no&yes&yes&no&no&no&yes\\
\hline
Dust evolution &no&no&no&no&yes&yes&yes&yes\\
&&&&&&&&no growth$^{1}$ \\
\hline
\hline
\end{tabular}
\tablefoot{{$^{1}$In this case all the particles are assumed to have the same size and only the transport is considered for the evolution.}}     
\label{assumed_models}
\end{table*}

The midplane temperature was calculated assuming energy equilibrium and contribution from optically thin and thick emission \citep{nakamoto1994}, as explained in \cite{birnstiel2010}, but without including viscous heating. The temperature profile approximates to a power law $T\propto r^{-1/2}$, such that at 1~au the temperature is $\sim130~$K, which are typical values for disks around T-Tauri stars.

For the dust evolution we followed the evolution of 180 grain sizes, and initially we assumed all grains to have the same size of 1~$\mu$m and a dust-to-gas ratio of 1/100 in the whole disk. The grain size grid covers up to 200~cm. The radial and grain size grids are logarithmically spaced. The fragmentation velocity of the particles is taken constant with a value of 10$\mathrm{m~s}^{-1}$ in the entire disk, as expected for grains with water ice mantles. This value is in agreement with the assumed disk temperature. Hence, in these models there is not radial changes of the fragmentation velocity of the particles as it is expected when the water snow-line is considered \citep[e.g.][]{birnstiel2010, banzatti2015}.  For the radiative transfer modelling, we assumed the disks to be face-on at a distance of $d=140~$pc. The stellar, disk, and dust parameters are summarised in Table~\ref{parameters}. 

In total, we investigated eight different models to study the gas and dust evolution when only viscous accretion is considered and when a dead zone of different shapes and/or an MHD wind is also assumed. The specific setup of each model is summarised in Table~\ref{assumed_models}.

\section{Results}     \label{results}

\subsection{Gas evolution}

\subsubsection{Dead zone included} Figure~\ref{model1_2} shows the gas surface density profile ($\Sigma_g$) at different times of evolution ($[0.1, 0.5, 1, 5]~$Myr) assuming a dead zone in the disk (model~\upperRomannumeral{2}, Table~\ref{assumed_models}). For comparison, the reference model that only assumes viscous accretion is also plotted (model~\upperRomannumeral{1}). In addition, $\alpha(r,t)$ (Eq.~\ref{dead_zone}) is also shown. At the location of the transition from a dead to an active region (the outer edge of a dead zone), an accumulation of gas forms due to the reduction of gas accretion in the dead zone ($\dot{M}\propto\nu\Sigma_g$). At long times of evolution (0.5 or 1~Myr), the amplitude of the gas bump becomes higher (a factor of 5 approximately) and in the outer part of the disk $\Sigma_g$ decreases steeply with radius.  The gas bump slightly moves inwards with time and after several million years of evolution ($\gtrsim$1~Myr), it spreads throughout the inner disk due to the viscous accretion. The lifetime of the bump is shorter ($\sim0.1~$Myr) when solving the hydrodynamical equations, instead of Eq.~\ref{eq_gas_evo}, which assumes Keplerian gas rotation. However, 2D/3D hydro- and MHD simulations have shown that the bump develops and lives longer with a similar amplitude as in Fig.~\ref{model1_2} \citep[$\sim10^{5}-10^{6}$~years,][]{matsumura2007, matsumura2009, regaly2012, flock2015}. We presume that the magnetic field configuration in combination with the Rossby wave instability (RWI) can mantain this bump structure for a long time. Therefore, the timescales for the evolution and dissipation of the bump in Fig.~\ref{model1_2} can be a reasonably estimate once the mentioned effects are taken into account (see Appendix~\ref{appendix_a}). However, it is important to clarify that solving Eq.\,\ref{eq_gas_evo} and assuming Eq.\,\ref{dead_zone} for $\alpha$ does not mimic fully 3D MRI simulations. Assuming viscous evolution is a simplification to study the dust evolution. 

The profile of $\alpha(r,t)$ shows how the outer edge of the dead zone becomes steeper and also moves inwards with time. This is because the gas bump also moves inwards and because in the regions where the gas density increases, the disk becomes less ionised (dead), and  thus $\alpha(r,t)$ decreases. Contrary, in the outer regions, the gas density is lower and it sharply decreases with radius, where the disk is expected to be more ionised (active).

\begin{figure}
 \centering
      	\includegraphics[width=9cm]{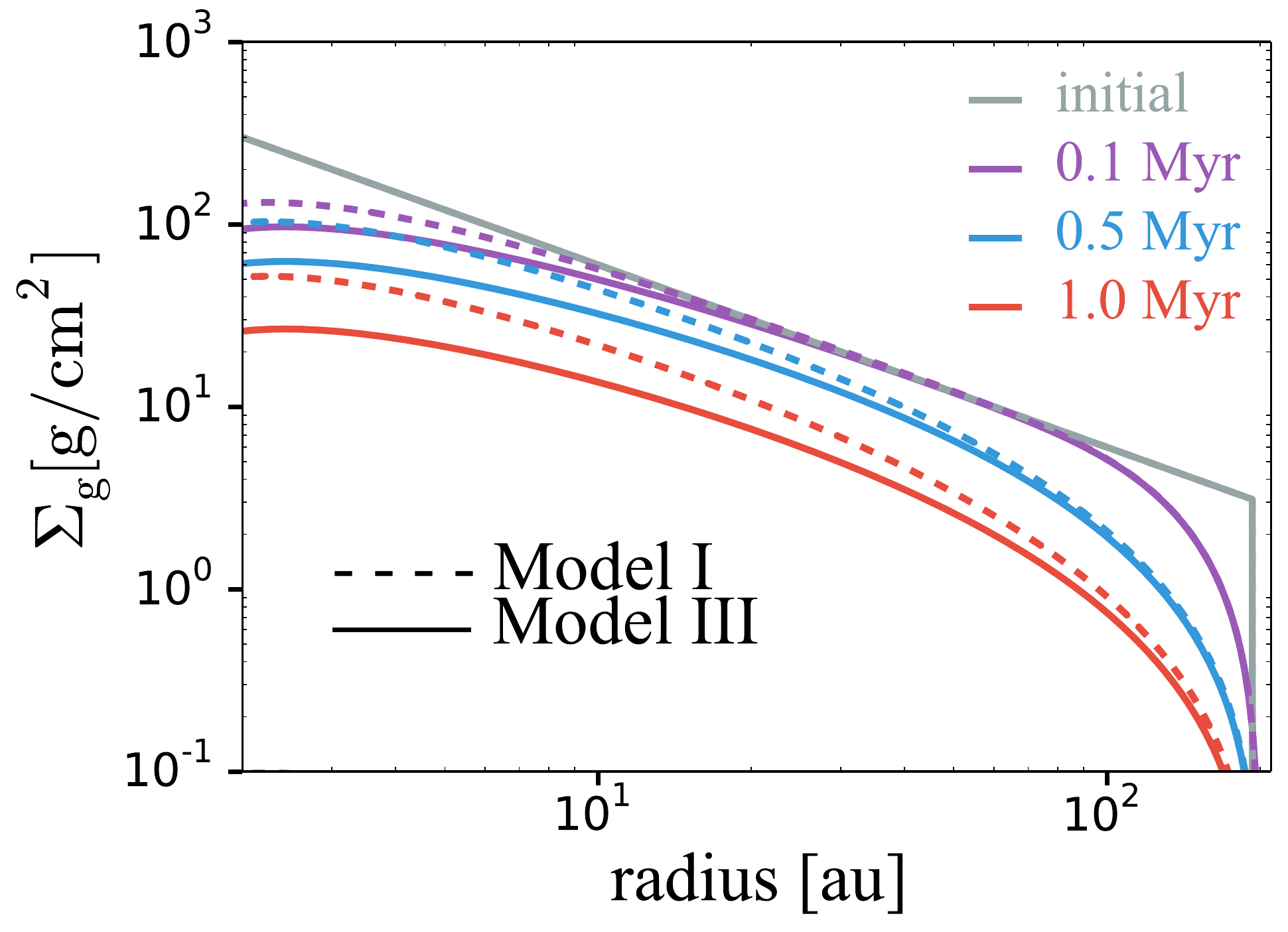}
    \caption{Effect of having an MHD wind ($C_w=2\times10^{-5}$, $\alpha=10^{-2}$, model~\upperRomannumeral{3}, solid lines) on the evolution of the gas surface density and the comparison with only viscous accretion (model~\upperRomannumeral{1}, dashed lines).}
   \label{model3}
\end{figure}

\subsubsection{MHD wind included} Figure~\ref{model3} shows the effect of having an MHD wind ($C_w=2\times10^{-5}$, $\alpha=10^{-2}$, model~\upperRomannumeral{3}) on the evolution of the gas surface density. In the inner part of the disk ($r\lesssim20$~au) and within the radial range of our simulation ($r>1$~au), the gas is slightly more depleted when a wind is included compared to the case where only viscous accretion is considered (model~\upperRomannumeral{1}), as previously found by \cite{ogihara2015}, who explored the effect of different wind strengths. For this work we maintained the value of $C_w=2\times10^{-5}$ constant for most of the simulations, which is a good representative for a weak net vertical magnetic field \citep{suzuki2010, ogihara2015}. With this value of $C_w$ and the assumed gas surface density (Eq.~\ref{eq:initial_gas}), the mass-loss rates are of the order of $\sim10^{-8}~M_\odot~$year$^{-1}$, which can be the case for strong MHD winds \citep{simon2013, gressel2015, bai2016b}.

\subsubsection{Dead zone and an MHD wind included}  
\begin{figure}
 \centering
      	\includegraphics[width=9cm]{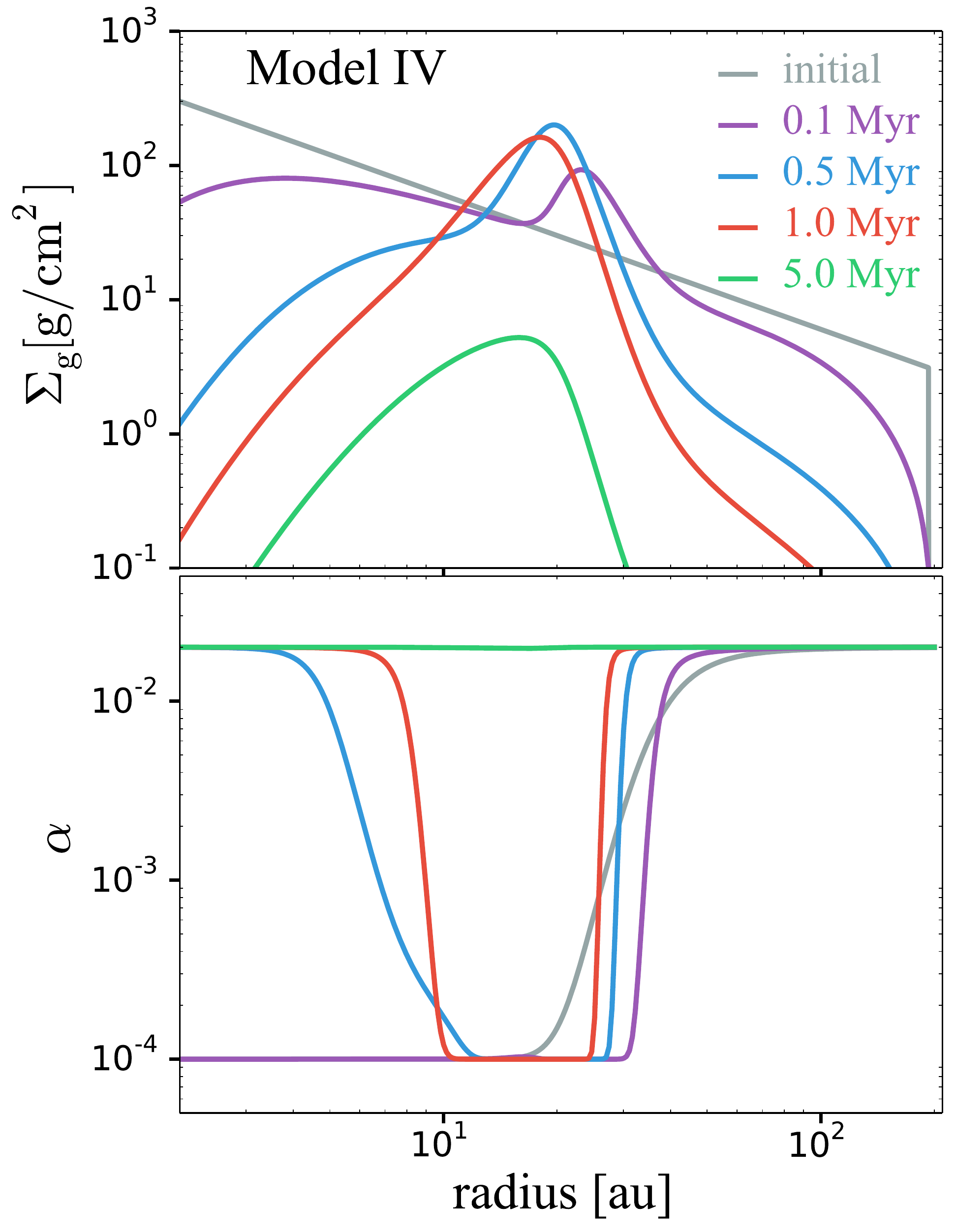}
    \caption{Evolution of the gas surface density (top panel) and $\alpha(r,t)$ profile (bottom panel) when a dead zone and an MHD wind is assumed in the disk  (model~\upperRomannumeral{4}).}
   \label{model4}
\end{figure}

Figure~\ref{model4} shows the case when a dead zone and an MHD wind are included for the evolution of the gas surface density (model~\upperRomannumeral{4}). A bump in the gas surface density is formed at the outer edge of the dead zone as in model~\upperRomannumeral{2} (Fig.~\ref{model1_2}). However, in this case the amount of gas efficiently decreases with time in the inner part of the disk, where it is dead ($r\lesssim10-20$~au) compared to the two previous cases. The inclusion of both phenomena allows forming a distinct, long-lived bump of gas close to the original outer edge of the dead zone. Because the inner part of the disk is depleted with time, it is expected  that the gas is ionised by UV, X-ray, or cosmic ray ionisation and therefore it should show high MRI activity again, as is reflected in the shape of $\alpha(r,t)$. At very long times of evolution ($\sim$5~Myr), $\alpha(r,t)$ is high and constant in the entire disk, but a strong and wide bump of gas remains. To create this robust ring in the gas surface density, the disk wind must have a stronger effect than the viscous accretion. As a test, we also assumed lower values of $C_w$ ($2\times10^{-7}$ and $2\times10^{-6}$), such that the ratio of $\alpha/C_w$ increased inside the dead zone. In these cases the gas surface density profile evolved in a similar way as in the case without a wind. 

Summarising the results for the gas evolution, when a dead zone is included, accretion slows down due to the radial changes of $\alpha$ within the dead zone, and a pile-up in the gas surface density profile is formed at the outer edge of the dead zone, which slowly spreads inwards. Since $\alpha$ depends on the gas surface density, when $\Sigma$ decreases, accretion reduces $\Sigma$ further and at long times of evolution the outer disk is depleted. When an MHD wind is included, it acts like accretion, mass disappears in the inner region, but these regions are refilled efficiently from the outer parts and there are no significant changes compared to the reference model (model~\upperRomannumeral{1}). When both dead zone and an MHD wind are included, it allows (together with the dependence of $\alpha$ on $\Sigma$) for an inside-out creation of a gas cavity, and a distinct bump in the gas surface density remains at long times of evolution. 

\subsection{Dust evolution}

\subsubsection{Dead zone included}  
\begin{figure*}
 \centering
      	\includegraphics[width=18cm]{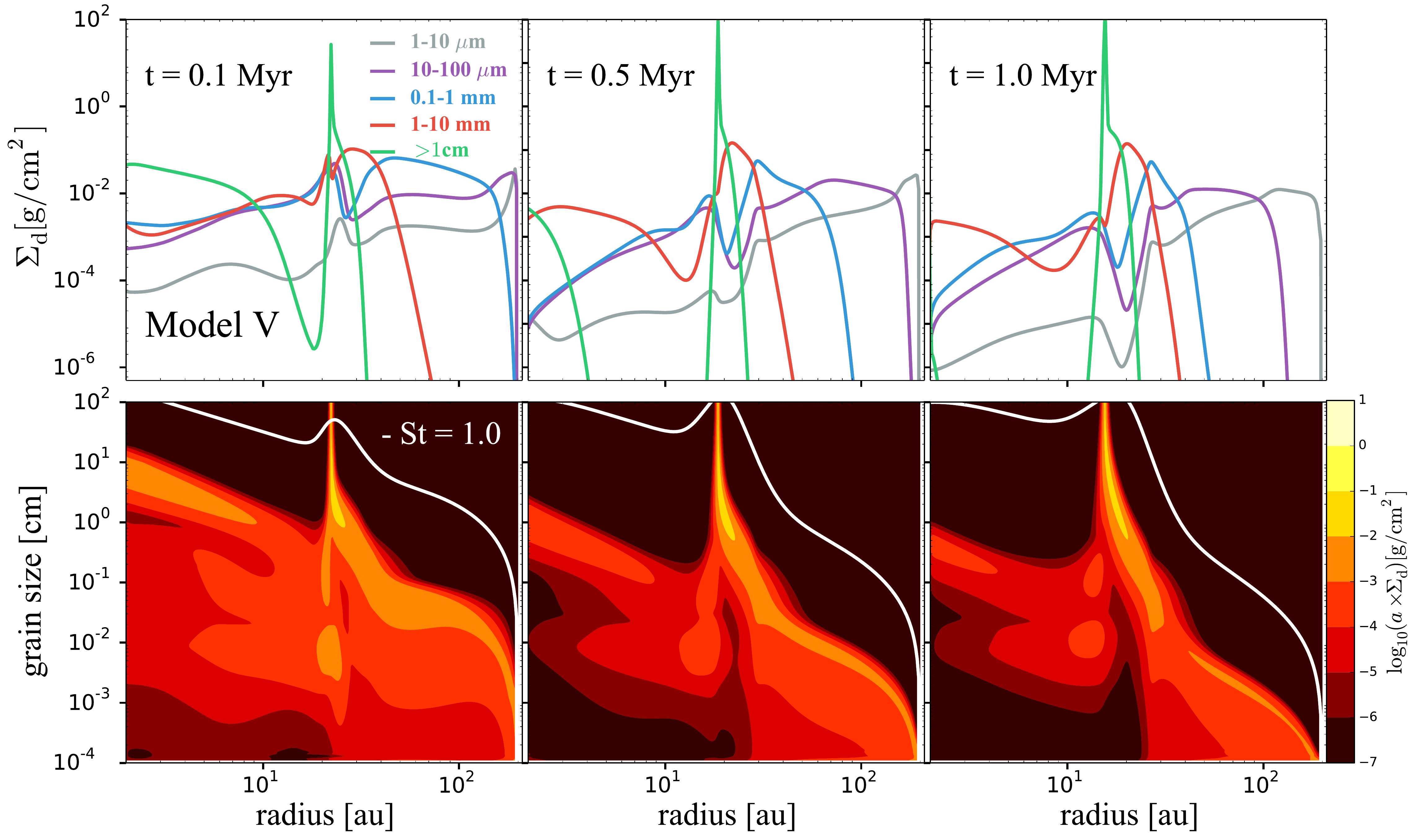}
    \caption{Dust density distribution of different dust species (top panels) and vertically integrated dust density distribution (bottom panels) at 0.1, 0.5, 1~Myr of evolution when a dead zone is included for the disk evolution (model~\upperRomannumeral{5}). The white solid line in the bottom panels corresponds to $\rm{St}=1$ (Eq.~\ref{eq:stokes}).}
   \label{model5}
\end{figure*}

Figure~\ref{model5} shows the dust density distribution of different dust species and the total dust distribution at three times of evolution ($[0.1, 0.5, 1]~$Myr) when a dead zone is assumed for the disk evolution and when the bump in the gas surface density profile still exits. In Fig~\ref{model5}, the $\rm{St}=1$ (Eq.~\ref{eq:stokes}) is also plotted, which is proportional to $\Sigma_g$ and it hence represents the gas surface density profiles shown in Fig~\ref{model1_2}. As a result of the gas bump that  is formed at the outer edge of the dead zone, there is a region of high density (hence high pressure), where particles reduce their radial drift velocity and move towards the pressure (or gas density) maximum. Particles with $\rm{St}\gtrsim\alpha$ effectively move towards pressure maxima and are trapped for longer times of evolution \citep[e.g.][]{brauer2008, pinilla2012, birnstiel2013}, which correspond to particles of $100-200~\mu$m in the dead-zone regions $r\lesssim 20$~au.  In the active zone, the particles with $\rm{St}\gtrsim\alpha$ correspond to sizes of around 1~mm. In the pressure bump, particles accumulate and efficiently grow since their relative velocities are low, which prevents fragmentation. In addition to radial drift, turbulent relative velocities are very low inside the dead zones, and this is the reason why particles can grow to very large sizes in the inner parts of the disk ($r\lesssim20$~au, Fig.~\ref{model5}), as similarly found by \cite{ciesla2007} at a fixed radius in the disk and \cite{brauer2008b} for the radial evolution. The effective growth causes most of the grains to be very large ($a\gtrsim$1~cm) inside the bump, depleting this region in small particles at long times of evolution (0.5-1~Myr) and creating gaps of small particles ($a\lesssim$1~mm) at the location of the gas bump. The shape of the ring-like accumulation of the millimetre-sized particles depends on the time of evolution. It becomes narrower in radial extent at longer times.

The timescales on which a pile-up of large grains inside the gas bump is achived can be estimated by considering the growth timescales to reach sizes of  $\rm{St}\gtrsim\alpha$ in addition to drift timescales, both at the location of the bump, that is, at $\sim20\,$au. On one hand, the growth timescales are inversely proportional to the local Keplerian frequency and the dust-to-gas ratio \citep{brauer2008}, that is, $\tau_{\rm{growth}}\propto \Omega^{-1}\Sigma_g /\Sigma_d$. At 20\,au, the growth timescale is around 1500 years ($\sim16\,$orbits at 20\,au, for a dust-to-gas ratio of 1\%). On the other hand, the drift timescales are given by \citep{birnstiel2012a}

\begin{equation}
	\tau_{\rm{drift}}=\frac{r^2\Omega}{\rm{St} c_s^2} \left | \frac{d\ln P}{d\ln r}\right  | ^{-1}.
 	\label{eq:t_drift}
\end{equation}

For particles to reach the drift size and accumulate in the bump, a certain number of growth times in addition to one drift time are necessary. Assuming our initial conditions, $\tau_{\rm{drift}}\sim12\,\tau_{\rm{growth}}$, which implies that the time for the particles to grow and to accumulate in the bump is approximately  $\sim24\tau_{\rm{growth}}$ or 36000\,years. This is a good approximation of the time that we obtain in our models, considering the transport of the dust by different mechanisms, the fragmentation, and the growth. Therefore, the gas bump created at the outer edge of the dead zone must live at least this time to form a ring-like accumulation of the large particles.

The dust trap survives as long as the gas bump exists at the outer edge of the dead zone. Therefore, at long times of evolution ($\gtrsim$1.0~Myr, Fig.~\ref{model1_2}), the dust trap disperses since the gas bump viscously smears out. When no traps are present in the disk, millimetre and centimetre particles are lost towards the central star after several million years of evolution \citep[see e.g. Fig.~4 from][]{pinilla2012}. \cite{birnstiel2012} investigated the dust evolution in a disk where the inner region has an increased surface density and a reduced $\alpha$, to mimic the effect of a dead zone in the disk without including the gas evolution. They concluded that under such conditions it is not possible to reproduce observations of disks with large mm cavities. In this paper, the combination of gas and dust evolution causes the trapping and creates dust-ring-like structures at the edge of a dead zone.

\begin{figure}
 \centering
      	\includegraphics[width=9cm]{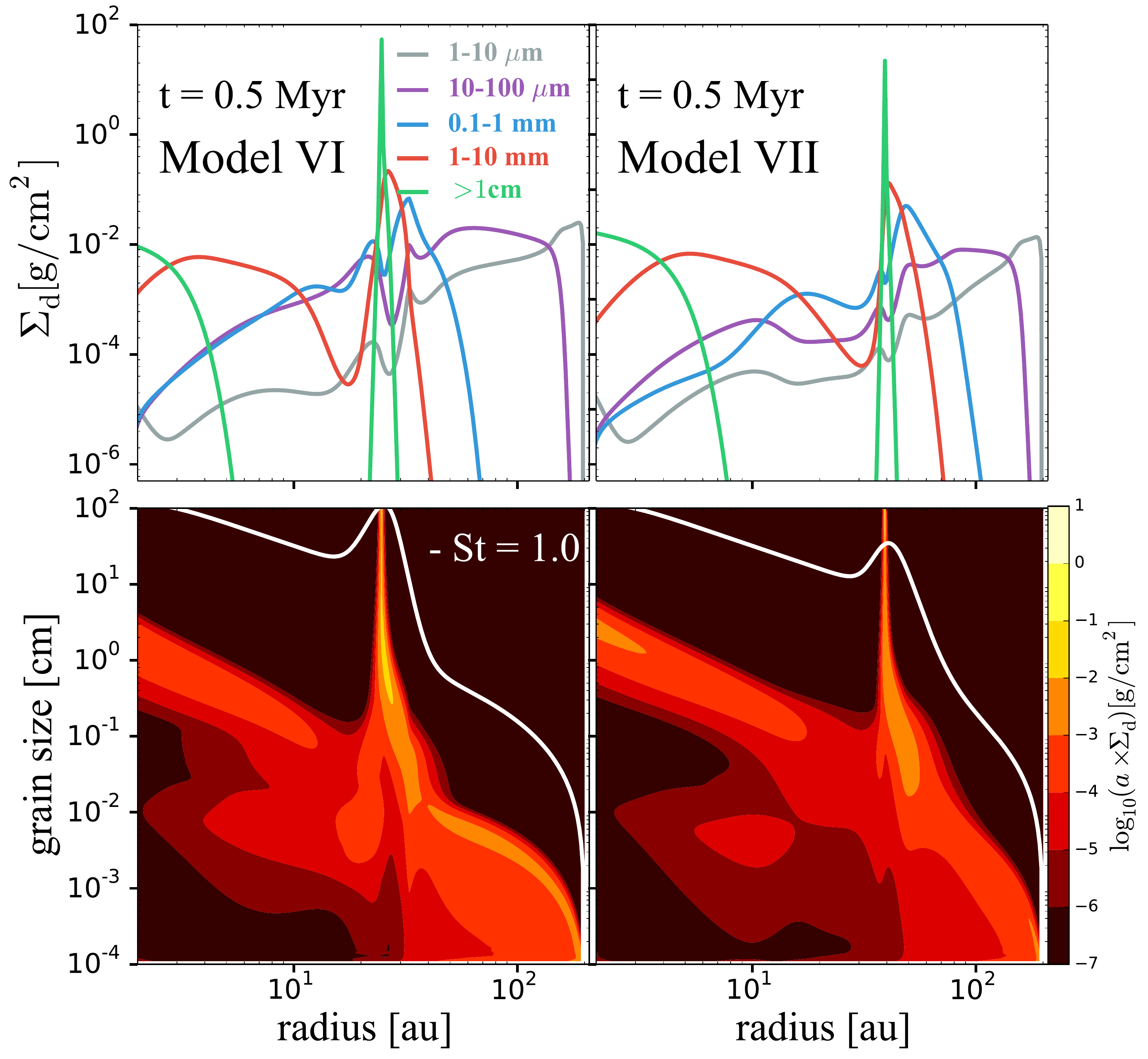}
    \caption{Dust density distribution of different dust species (top panels) and vertically integrated dust density distribution (bottom panels) at 0.5~Myr of evolution for a steeper dead zone edge (left panels, model~\upperRomannumeral{6}) and when the dead-zone edge is located farther out (right panels, model~\upperRomannumeral{7}). The white solid line in the bottom panels corresponds to $\rm{St}=1$ (Eq.~\ref{eq:stokes}).}
   \label{model6_7}
\end{figure}

Based on the diversity of the dead-zone morphologies found for different disk properties \citep[e.g.][]{dzyurkevich2013}, we investigated two more cases where (i) the transition from the dead to the active zone is steeper and occurs from $\sim$25-40~au compared to 20-60~au, and (ii) when the transition occurs farther out at $\sim40~$au (Table~\ref{assumed_models}). Figure~\ref{model6_7} shows the dust density distribution at 0.5~Myr of evolution for these two cases (models~\upperRomannumeral{5} and \upperRomannumeral{6}). In these cases, the results are similar to those in Fig.~\ref{model5}, that is, an effective trapping and growth of particles at the edge of the dead zone and a decrement of small particles at that location ($a\lesssim$1~mm).  By comparing the same time of evolution, the accumulation of large particles is narrower in the case with a steeper dead-zone edge. For model~\upperRomannumeral{5}, the centimetre-sized or larger particles accumulate in a ring of $\sim10$~au width and millimetre-sized grains are located in a ring of $\sim$30~au width, which are about half of the width for the steeper case. The dust surface density of the millimetre-sized particles is  higher inside the trap for the steeper dead-zone edge since the pressure gradient is higher and therefore the trapping of particles with  $\rm{St}=1$ is more effective. Because the pressure gradient becomes higher with time than in model~\upperRomannumeral{4}, the drift timescales are shorter (Eq.~\ref{eq:t_drift}), and as a consequence, the time to achieve a pile-up of millimetre-particles is also shorter (28000\,years, considering all the mechanisms of grain growth, fragmentation, and transport).

\begin{figure*}
 \centering
      	\includegraphics[width=18cm]{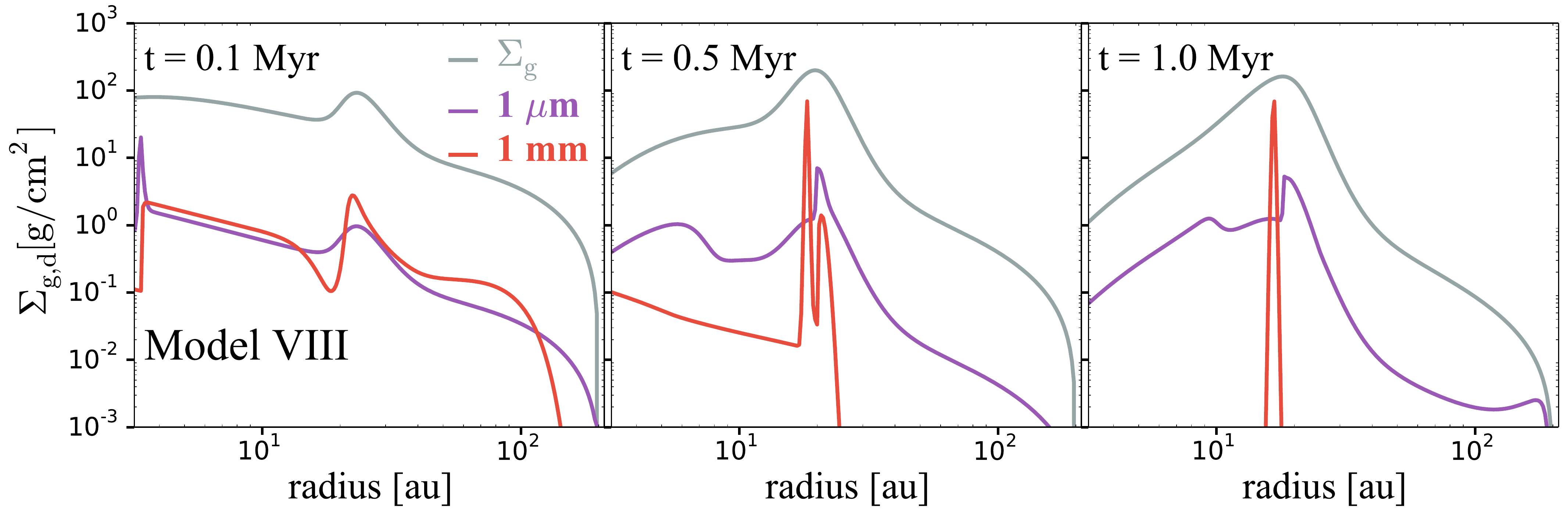}
    \caption{Evolution of the dust surface density distribution of a single grain size ($1\mu$m and 1~mm) assuming a dead zone and an MHD wind in the disk (model~\upperRomannumeral{8}). In this model all the particles are assumed to have a single size and only the transport is considered for the evolution.}
   \label{model8}
\end{figure*}

In the case that the dead zone edge is located farther out, the results are also similar as in Fig.~\ref{model5}. The main difference is that the density for large grains ($a\gtrsim$1~cm) is lower in this case. This is because the largest size that particles can reach before they fragment is smaller in the outer part of the disks \citep[$a_{\rm{frag}}\propto\Sigma_g$, e.g.][]{birnstiel2010}, and therefore there is a smaller reservoir of large particles to be trapped and grow ($a_{\rm{frag}}$ is close to the particle size that is efficiently trapped, i.e. when $\rm{St}\sim\alpha$).  In this case, the growth timescales are longer (Eq.~\ref{eq:t_drift}), and it takes about 60000\,years to start piling up millimetre-sized particles at the outer edge of the dead zone.

\subsubsection{Dead zone and an MHD wind included} \label{sect_wind}
For the calculation of the dust evolution in the case where a dead zone and an MHD wind are both included, some simplifications were made. At the inner boundary the calculation of  the growth and the transport of the dust particles is numerically challenging because of the effective removal of gas in the inner part of the disk (Fig.~\ref{model4}). As a simplification and to illustrate how the dust distribution is expected to be in this case, we investigated the evolution and transport of dust when all particles had the same size and we did not include the growth. 

Figure~\ref{model8} shows the evolution of the dust surface density distribution of $1\,\mu$m- and 1\,mm particles assuming a dead zone and an MHD wind in the disk (model~\upperRomannumeral{8}). These two grain sizes are representative of the behaviour of small grains that are coupled to the gas and large grains that become trapped in the gas bump.  These results illustrate how the large grains are concentrated in a narrow ring at the locations of the gas maximum while small grains are not being trapped but follow the gas. Inside the dead zone, fragmentation is expected to be again efficient as a result of the reduction of the gas surface density, even when turbulent velocities are low. The inner part of the disk is expected to be depleted of large particles ($a\gtrsim1$~mm) and form gap-like low-emission regions, where gas is also depleted, but by a lower factor. The depletion of the gas and large particles strongly varies with time and at 1~Myr of evolution, the gas depletion reaches about three orders of magnitude while the millimetre dust is depleted by more than four orders of magnitude. In this case, a radial segregation between the gap in the gas and the millimetre-sized particles is expected.

\subsection{Synthetic images and radial profile emission}

\begin{figure*}
 \centering
      	\includegraphics[width=18cm]{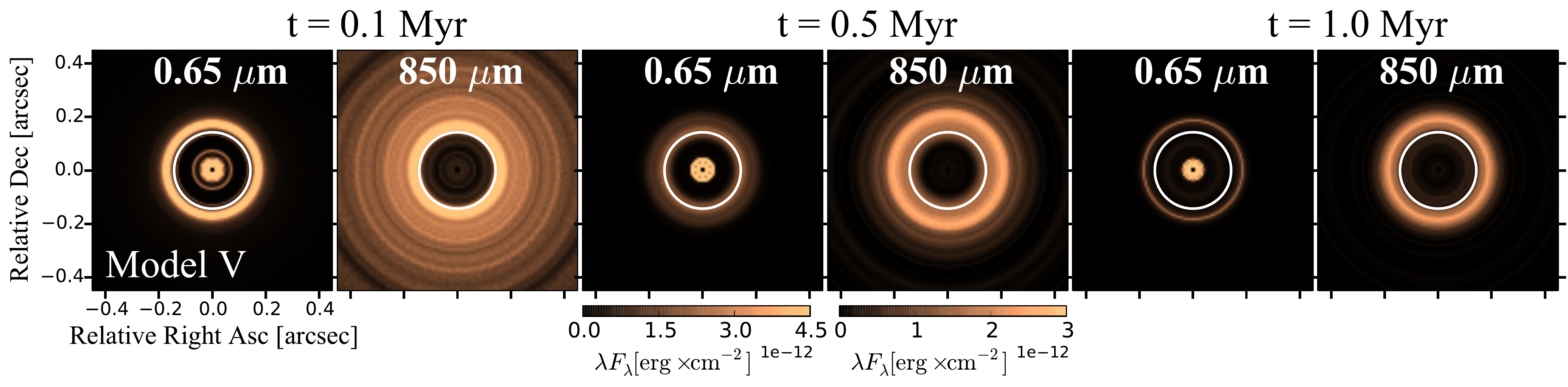}\\
	\includegraphics[width=18cm]{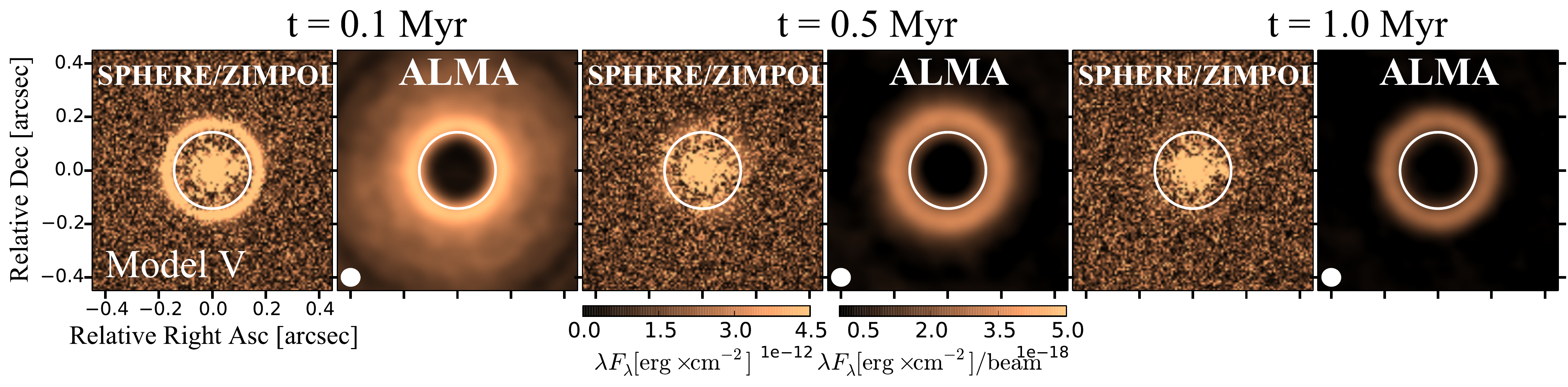}
    \caption{Synthetic images at $0.65\mu$m (polarised intensity) and at $850\mu$m obtained after radiative transfer calculations, before and after instrument simulations (SPHERE/ZIMPOL and ALMA), using the dust density distributions obtained from model~\upperRomannumeral{5} (Fig.~\ref{model5}, dead zone alone) at different times of evolution. Circles at 20~au (assuming 140~pc distance) is also shown for reference. The star is not suppressed for the  scattered optical light images. }
   \label{images_model5}
\end{figure*}

To show realistic images for some specific cases, we also performed instrument simulations at $0.65~\mu$m and at $850~\mu$m wavelengths from the model images obtain after the radiative transfer calculations. We used the Common Astronomy Software Applications (CASA \footnote{\url{http://casa.nrao.edu/}}), to create ALMA simulated images in Band~7 ($\sim850~\mu$m), including atmospheric noise and assuming an antenna configuration that allows a final resolution of $0.05''$. We assumed an hour for the total observing time. At R band (0.65$\mu$m) we used the SPHERE simulator \citep{thalmann2008}, which includes realistic resolution, sensitivity, flux loss, etc. For more details of similar observational simulations, we refer to \cite{dejuanovelar2013}.

Figure~\ref{images_model5} shows the synthetic images before and after instrument simulations, for the polarised emission at $0.65~\mu$m and the thermal emission at 850~$\mu$m. For the case of a dead zone alone, we show three different times of evolution ($[0.1, 0.5, 1]~$Myr) obtained after the radiative transfer calculations, for which the dust density distributions from model~\upperRomannumeral{5} (Fig.~\ref{model5}) are assumed. These images show that at early times of evolution (0.1~Myr) a ring-like emission at $0.65~\mu$m exists at around $\sim20$~au and a gap of similar size is formed at millimetre-emission, where the gas bump is located (Fig~\ref{model1_2}). As a result of the effective growth at those locations, the small particles (all grains with $a\lesssim100~\mu$m) are depleted, forming a gap (Fig.~\ref{model5}), and therefore the ring at the $0.65~\mu$m-polarimetric images that exists at 0.1~Myr vanishes at longer times of evolution (i.e. 0.5 and 1~Myr). Instead, a narrower and very faint ring remains farther out at $\sim30$~au, where the outer edge of the gap in small grains is located. This is also the location where the disk scale height is expected to become higher since here $\alpha(r,t)$ increases (Fig.~\ref{model1_2}), and the small grains can be vertically distributed by higher turbulence (in contrast to the dead zone, where the disk is expected to be flatter). A similar result of a directly irradiated dust wall at the outer dead-zone edge was found by \cite{hasegawa2010}. However, these rings at $0.65~\mu$m smear out after the instrument simulation where only the stellar emission is significant (see SPHERE/ZIMPOL image in Fig.~\ref{images_model5} at 0.5 and 1~Myr of evolution). 

In this case, the ring at millimetre-emission becomes narrower because of radial drift towards the pressure maximum, but it also becomes fainter because of the effective growth to larger grains ($a\gtrsim$1~cm), which have very low opacities and hence lower millimetre fluxes. The structures from Fig.~\ref{images_model5} remain as long as the gas bump exists ($\sim1$\,Myr). It is important to note that the rings at optical wavelengths are located either at a similar distance as the mm rings or slightly farther out. 

\begin{figure*}
 \centering
 \tabcolsep=0.1cm 
   \begin{tabular}{cc}   
   	\includegraphics[width=6cm]{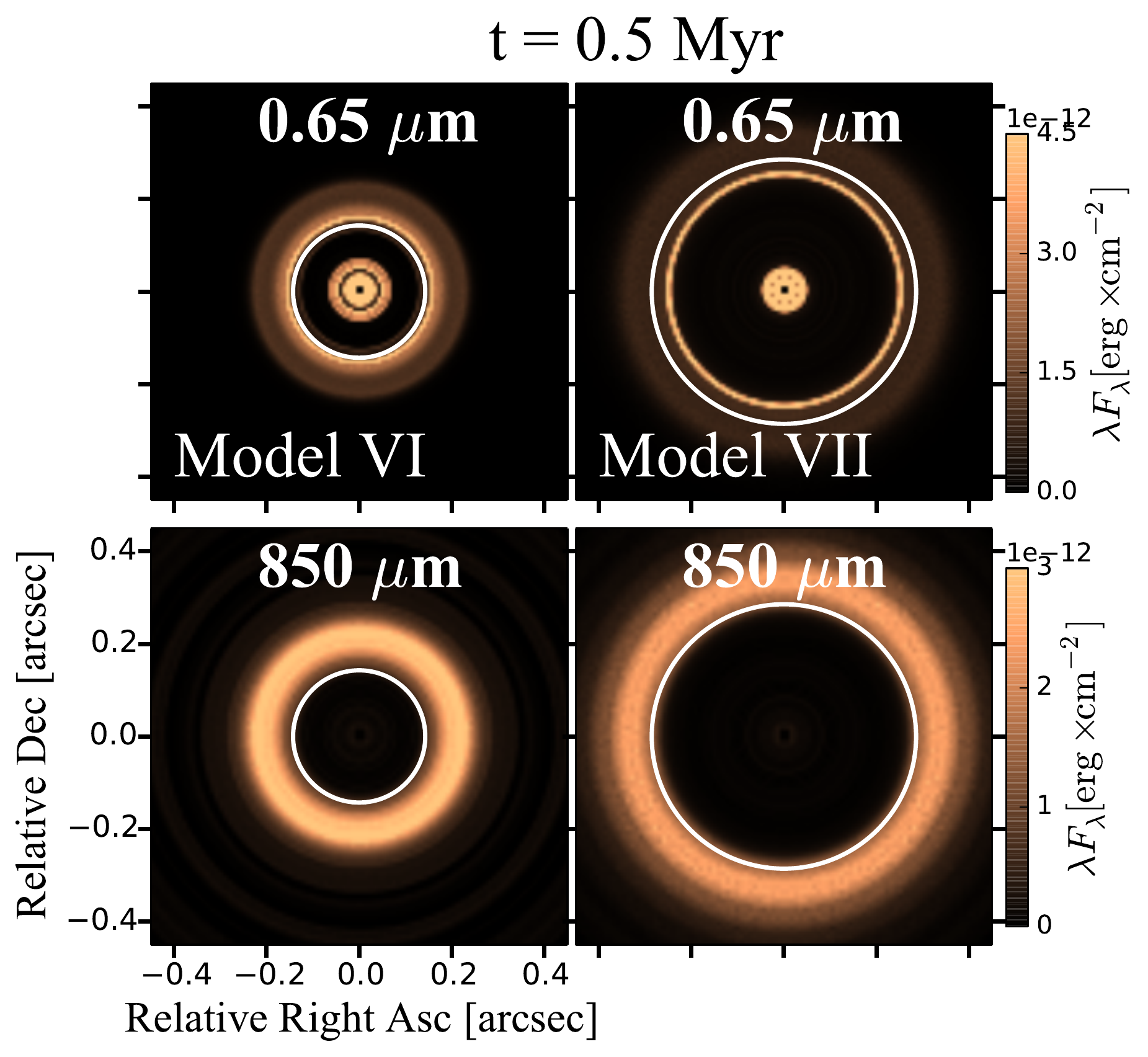}&\includegraphics[width=12cm]{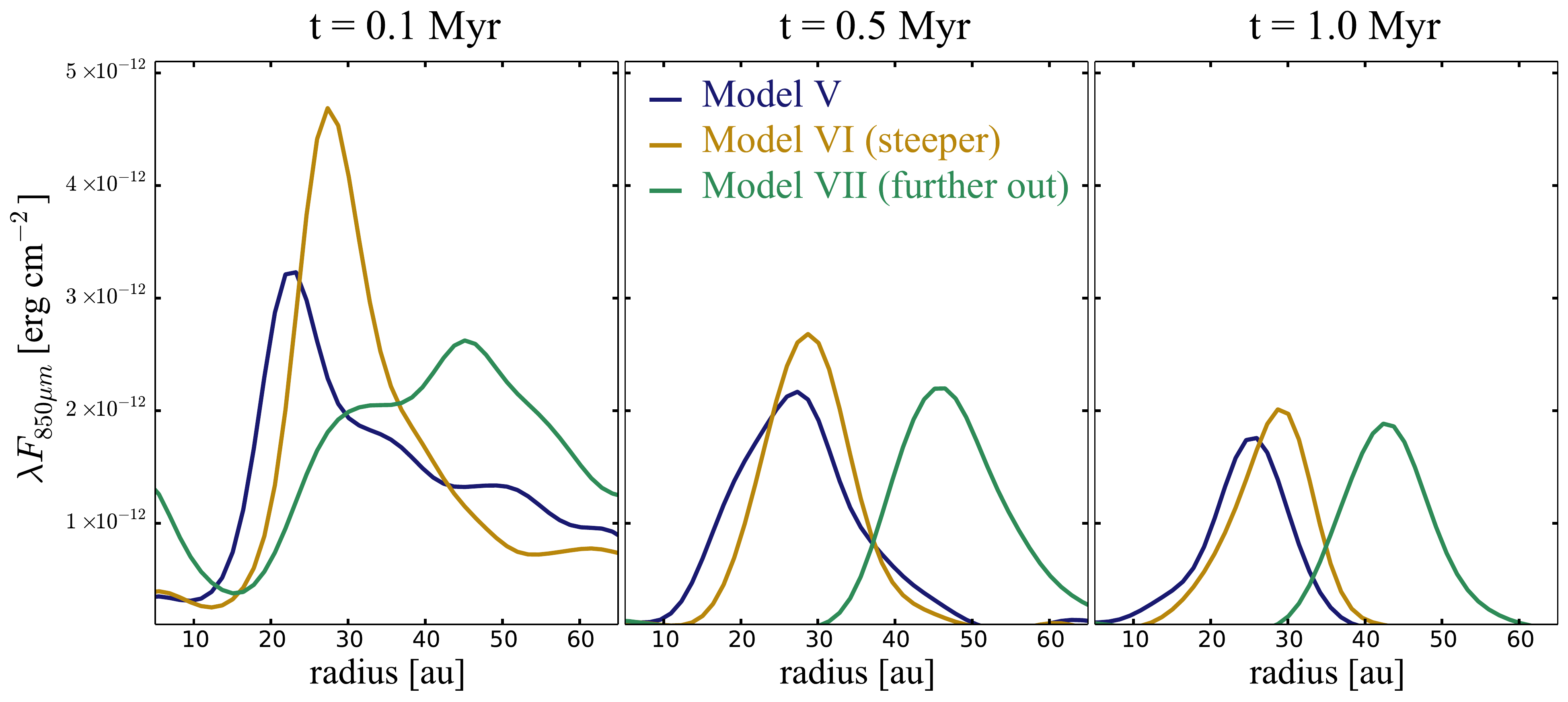}
   \end{tabular}
    \caption{Left panel: synthetic images at $0.65\mu$m (polarised intensity, left panels) and at $850\mu$m  (right panels) obtained after radiative transfer calculations using the dust density distributions obtained from models~\upperRomannumeral{6} and  \upperRomannumeral{7}  (Fig.~\ref{model6_7}) at 0.5 Myr of evolution. Circles at 20~au in the left  and 40~au in the right (assuming 140~pc distance) are also shown for reference. Right panel: radial intensity profile at 850~$\mu$m from the model images of models~\upperRomannumeral{5}, \upperRomannumeral{6}, and  \upperRomannumeral{7} at  different times of evolution ($[0.1, 0.5, 1]~$Myr).}
   \label{images_model6_7}
\end{figure*}

The left panel in Fig.~\ref{images_model6_7} shows the synthetic images for the polarised emission at $0.65~\mu$m and the thermal emission at 850~$\mu$m emission for models~\upperRomannumeral{6} and \upperRomannumeral{7} at 0.5~Myr of evolution. For the case when the dead-zone edge is steeper, the structures are similar as in model~\upperRomannumeral{5}, but they are brighter and narrower, in agreement with the moderately higher density of the mm grains and the steeper increment of the dust surface density of the small grains (1-10~$\mu$m) at the outer edge of the gap. For the dead-zone edge that lies farther out, the structures are similar, but the $0.65~\mu$m-polarised image boosts the inner edge of the formed gap in small grains, since in this model there are marginally more small grains at farther locations because growth occurs at a slower rate than in the inner parts of the disks. As a consequence, in this case the $0.65~\mu$m-ring is slightly closer than the mm ring. This mixing of small grains at the gap just in front of the pressure maximum seems robust and was also observed in global 3D non-ideal MHD simulations that included particles of different sizes \citep[see Fig. 3 in][]{ruge2016}.

The radial intensity profile at 850~$\mu$m from the synthetic images of models~\upperRomannumeral{5}, \upperRomannumeral{6}, and  \upperRomannumeral{7} at  different times of evolution ($[0.1, 0.5, 1]~$Myr) is shown in the right panel in Fig.~\ref{images_model6_7}. These profiles confirm that at 0.1~Myr of evolution the ring at mm emission is narrower and brighter when the dead-zone edge is steeper. At 0.5~Myr, the mm ring has a similar brightness for the three cases, but the total width of the mm ring remains smaller for the  case of a steeper edge of the dead zone ($\sim$20~au for model~\upperRomannumeral{6} vs. $\sim$30~au for models~\upperRomannumeral{5} and  \upperRomannumeral{7}). At 1.0~Myr, the mm ring is similar for the three cases. For each time of evolution, the peak of the mm emission is slightly farther out when the dead-zone edge is steeper.   

We emphasise that we dis not include synthetic images of the models that take into account the effect of an MHD wind for two different reasons. First, we did not perform dust evolution models that include the growth for this case (Sect.~\ref{sect_wind}) and therefore the resulting synthetic images would not be comparable with those shown in Figs.~\ref{images_model5} and \ref{images_model6_7}. Second, the micron-sized particles can be carried out  upward by an MHD wind \citep[e.g.][]{miyake2016}, which changes the expected observable signatures at optical emission. However, the amount of small dust in the wind would depend on the launching region. If this occurs in the upper layers of the disk \citep[$>2-3h$ e.g.,][]{bai2016}, we expect that inside the dead zone not many small particles are left, which should be dragged by the wind due to the low turbulent mixing in this region. As was shown by \cite{zsom2011} (their Fig. 6), who used the same mixing parameter as we do in the dead zone,  nearly no grains survive at these upper layers. We therefore expect that the number of small grains dragged by the wind is small.

\section{Discussion}     \label{discussion}

We here investigate the simultaneous evolution of the gas and dust when a dead zone and/or an MHD wind act on the disk. When only a dead zone is assumed in the dust and gas evolution calculations, cavities and ring-like structures are formed at short and long wavelengths (Figs.~\ref{images_model5} and \ref {images_model6_7}), but the obtained structures are quite variable with time. At early times of evolution ($\sim0.1$~Myr), a ring is formed at a similar location at $0.65~\mu$m as at $850~\mu$m emission, which would be visible with instruments as SPHERE/ZIMPOL and ALMA. However, at these early times, it is expected that the disk is still surrounded by an envelope, which would prevent us from detecting the disk at optical or IR observations. At longer times of evolution (0.5 and 1~Myr), the ring at $0.65~\mu$m becomes very faint as a result of the effective growth of particles within the dead zone, and only the ring at mm emission remains bright for longer times ($\sim1$~Myr). The millimetre-sized particles at the outer edge of the dead zone pile up on shorter time scales than the expected lifetime of the gas bump, and the mm ring survives as long as the gas bump exists (see also Appendix~\ref{appendix_a}). 

Most of the observations of transition disks show a continuous distribution of dust or smaller cavities at optical or near-infrared (NIR) polarimetric images, but a clear cavity at the millimetre range \citep[e.g.][]{dong2012}. Some examples have been reported, however, where the scattered-light cavity is similar in size to the mm cavity found in our results of dead zones, such as IRS~48 \citep{follette2015}, HD~142527 \citep{avenhaus2014}, and LkCa~15 \citep{thalmann2015}. Nonetheless, in the case of LkCa~15, several planet candidates have been observed inside the mm cavity and at least one has recently been  confirmed \citep{sallum2015}. In HD~142527 a very massive companion at $\sim12$~au has been detected \citep{biller2012, close2014}, but it is still unclear whether this companion is responsible for the very wide dust cavity ($\sim140$~au). For IRS~48 and HD~142527, high-contrast asymmetries at mm emission have also been observed, which have been interpreted as long-lived vortices that trap particles. These vortices can originate from RWI, which can be triggered by massive planets \citep[e.g.][]{zhu2014}, but also at the edge of a dead zone \citep[e.g.][]{flock2015, lyra2015, ruge2016}. 

\begin{figure*}
 \centering
 \begin{tabular}{cc}  
      	\includegraphics[width=9cm]{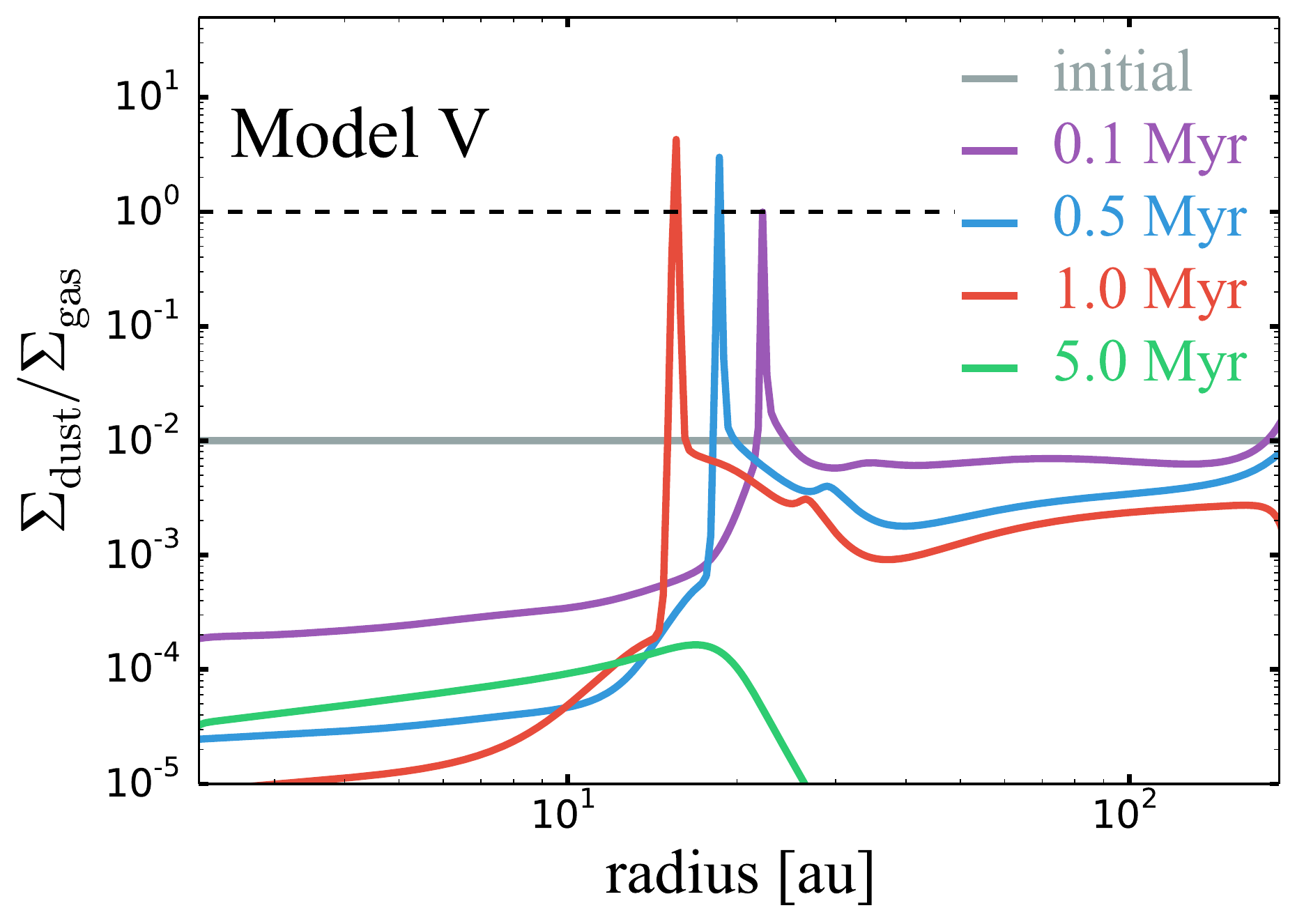}&\includegraphics[width=9cm]{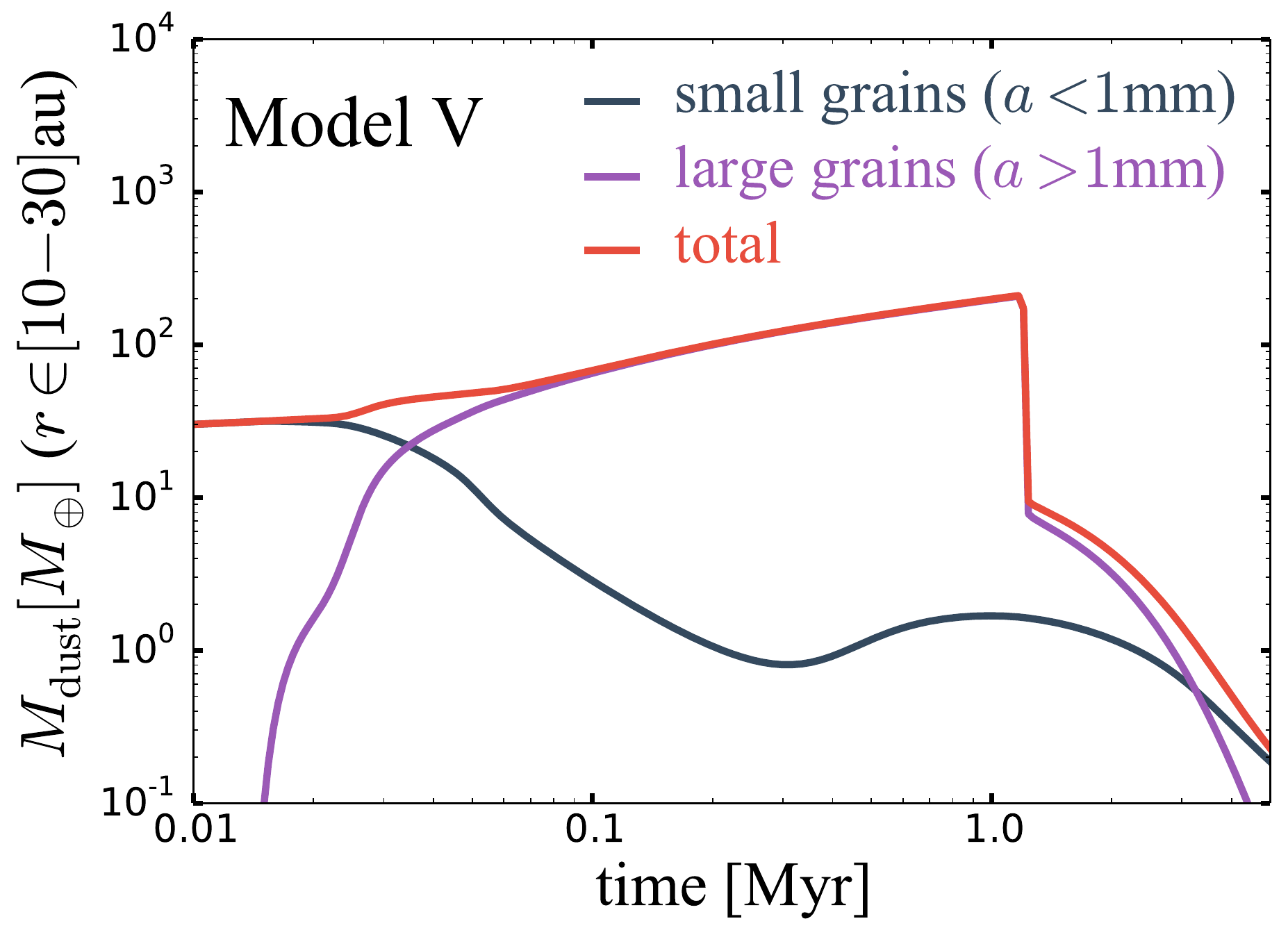}
  \end{tabular}
    \caption{Left panel: dust-to-gas surface density ratio for different times of evolution of model~\upperRomannumeral{5}.  Right panel: mass of dust as a function of time calculated between 10 and 30~au (within the region where the dust trap exists and moves during the simulation) for model~\upperRomannumeral{5}.}
   \label{masses}
\end{figure*}

The accumulation and growth of particles at the outer edge of the dead zone are effective enough to increase the dust-to-gas surface ratio and reach values exceeding unity (Fig.~\ref{masses}, left panel). In these cases, the feedback from the dust to the gas can lead to streaming instabilities and the fast formation of planetesimals \citep{youdin2005}. The mass of the accumulated dust in the region where a dust trap exists and moves during the simulation ($\sim10-30$~au) can reach values exceeding 100~$M_{\bigoplus}$ (which can be lower if the pressure bump lives for shorter times and/or the initial dust mass is lower), assuming an initial disk mass of $\sim0.08~M_{\odot}$ and a constant dust-to-gas ratio of 1/100 (Fig.~\ref{masses}, right panel). Recently, the 7\,mm observation of HL Tau allowed a more accurate estimation of the dust mass inside one of the outer rings, with estimates between 70 and 210~$M_{\bigoplus}$ \citep{carrasco2016}. In combination with our results this indicates that the pressure bump has to be longer lived (at least $\sim$0.03\,Myr, Fig.~\ref{masses}) to accumulate such highly solid material. If this mass can be assembled into a core, then it is sufficient to start gas accretion and form giant planets \citep[e.g.][]{pollack1996}.  Our calculations did not consider the feedback from the dust to the gas, and further work is required to investigate whether a massive planet can be formed at this location and trigger the formation of other pressure traps.  

Figure~\ref{masses} (right panel) also provides a clear evidence of the effective drift and loss of dust towards the star once the bump in gas viscously disperses at longer times of evolution ($\sim1.0$~Myr). However, photometry and spectroscopy data from \emph{Spitzer} suggest that the transition-disk frequency increases from $\sim15-20$\% at 1-2~Myr to 50\% at 5-8 Myr \citep{currie2011}. In the context of our models, this would imply that if the appearance of a transition disk is caused by a dead zone, then the gas bump formed at the outer edge should live longer. This can be  the case in more realistic 2D/3D hydro- and MHD simulations (Appendix~\ref{appendix_a}), and hence dead zones may be the origin of transition disks that are  $\gtrsim1\,$Myr in age as well.

In the scenario of a dead zone alone, the gas surface density bump at the outer dead-zone edge is increased by a factor of  5 compared to the surface density within the dead zone. This contradicts the gas depletion inside mm cavities of transition disks, which seems to be much higher \citep[e.g. $\delta_{\rm{gas}}=10^{-4}-10^{-2}$][]{marel2016}. The inclusion of an MHD wind in the simulations fosters the reduction of the gas in the inner regions of disks ($r\lesssim10-20$~au, Fig.~\ref{model3}), creating  gas depletion of about three orders of magnitude at 1~Myr of evolution (Fig.~\ref{model4}). This model also allows a ring shape for the gas that lives longer (up to $\sim5$~Myr) than when a dead zone alone is considered. When the dead zone edge is assumed to be located farther out, the ring is formed at later times of evolution.  Under our simple assumption that the disk wind efficiency is independent of turbulent viscosity ($\alpha$), the disk mass decreases significantly from $\sim0.08~M_{\odot}$ (initial condition) to $\sim0.02$ and $0.001~M_{\odot}$ at 1 and 5~Myr of evolution, respectively, for model~\upperRomannumeral{4} with $C_w=2\times10^{-5}$. The combination of these low values for the gas surface density and low values of $\alpha$ inside the dead zone would lead to very low disk accretion rates of $\lesssim10^{-11}~M_{\odot}\rm{year}^{-1}$. This value for the accretion rate is too low compared to observations of transition disks \citep[e.g.][]{manara2014, owen2015}, even though when an MHD wind is included, the disk is active at the end of the simulation. Nonetheless, MHD winds are expected to carry away angular momentum and drive accretion. Therefore the absolute value of $\alpha$ should increase when the disk-wind stress term is included. A more detailed investigation into the mutual dependence of $\alpha$ and into the strength of the disk wind is needed to obtain proper disk masses and accretion rates. In addition, within the dynamical range of our simulations, we also neglected the effect of an inner edge of the dead zone located closer-in, where thermal ionisation can increase turbulence.

\section{Summary}     \label{summary}

We studied the radial evolution of gas and dust in protoplanetary disks in the presence of a dead zone and under the influence of mass-loss caused by a disk wind. We assumed an $\alpha$-viscosity that depends on the surface density to mimic the effect of MRI-driven turbulence. Our main findings are summarised below.

\begin{itemize}
\item When a dead zone is assumed for the viscous evolution, a gas surface density bump is formed at the outer edge of the dead zone, which has a small amplitude (a factor of $\sim5$). Our results show a bump lifetime of more than 1\,Myr.  The dependence of $\alpha$ on the gas surface density leads to a depleted outer region (Fig.~\ref{model1_2}). The resulting formation of a gas bump causes a region of high pressure that is capable of trappin particles;  grain growth is also very efficient there (Fig.~\ref{model5}), and the dust-to-gas ratio can reach values above unity (Fig.~\ref{masses}). If the accumulated dust (which mass is  10-100~$M_{\bigoplus}$) forms a core, then it can be enough to start gas accretion and form giant planets. At longer times of evolution ($\gtrsim1.0~$Myr), the bump smears out and the dust trap vanishes.

\item The dust density distributions at timescales of several million years are not strongly dependent on the location of the outer dead-zone edge or on the steepness of the transition between dead and active zones (Fig.~\ref{model6_7}).

\item Synthetic images derived from radiative transfer simulations show that for the cases that include a dead zone, a cavity and a ring-like emission exist independent of the shape of the dead zone that it is assumed (Figs.~\ref{images_model5} and \ref{images_model6_7}). These structures strongly depend on the time of evolution, and the ring at 0.65~$\mu$m is bright and can be detected at early times ($\sim0.1$~Myr). The resulting structures are similar  to those created by planet-disk interaction processes \citep[for $\lesssim1~M_{\rm{Jup}}$ planets, e.g.][]{dejuanovelar2013, dong2015, gonzalez2015}. Nonetheless, none of the explored cases that only included a dead zone showed a continuous distribution of dust or smaller cavities at optical or NIR-polarimetric images while a clear cavity at the millimetre range, as is observed for different transition disks \citep[e.g.][]{follette2013}. 

\item When viscous accretion and an MHD wind are assumed for the gas evolution, the wind does not have a significant effect on the final gas surface density (with $\alpha=10^{-2}$ and $C_w=2\times10^{-5}$, Fig.~\ref{model3}). Nevertheless, when the same MHD wind and a dead zone are included, the gas in the inner parts of the disk ($r\lesssim10-20$~au, within the dead region) becomes highly depleted, and the final gas surface density shows a distinct wide ring (Fig.~\ref{model4}). The dead zone slows the accretion down and during that time the high mass-loss rate by the wind can have a much stronger effect in the inner region.  As a consequence of the gas depletion, the inner parts of the disks become active and at long times of evolution ($\sim5~$Myr), the dead zone disappears.

\item When a dead zone and an MHD wind are included, large particles ($a\gtrsim$1~mm) are expected to be concentrated in the peak of the gas density, which creates a large radial difference between the inner edge of the gas bump and the mm dust (Fig.~\ref{model8}). 
\end{itemize}

\section{Conclusion and outlook}     \label{conclusion}

The gas and dust structures that we observe when we assume a dead zone acts on the disk reproduce structures that are observed in transition disks. At the outer edge of a dead zone, a bump in the gas surface density forms where particles are trapped. The exact lifetime of the bump remains open, while it could be shorter in 1D hydrodynamical simulations it could be even longer in 3D non-ideal MHD simulations where the bump is sustained by MRI and RWI activity \citep[see Appendix~\ref{appendix_a} and][]{regaly2012, hasegawa2015, flock2015, miranda2016}. In the cases we explored, the gas bump lived long enough for the particles to grow, accumulate, and form emission resembling a ring at the mm range at the location of the gas bump.

In the synthetic images, a ring-like structure in scattered optical  light and millimetre continuum emission was obtained from the models by assuming a dead zone. The radial distance of the ring is similar at early times of evolution ($\sim$0.1~Myr) for both wavelengths. However, at longer times of evolution  ($\sim$0.5-1~Myr), no ring is seen at scattered light because of the effective growth of particles inside the trap that is formed at the outer edge of the dead zone. 

The gas within the dead zone is depleted by a small factor with respect to the gas bump, but this effect can be strongly enhanced by the presence of an MHD wind. Both phenomena (a dead zone and an MHD wind)
 can create a factor for the gas depletion similar to observations \citep[the gas surface density inside the mm-cavities can be depleted by a factor of 1000 or more, e.g.][]{marel2016}. This scenario is therefore a good candidate for mimicking the fact that the gas gap size is smaller and less depleted than the mm dust emission. Dead zones can also reproduce other asymmetric structures such as vortices \citep[e.g.][]{flock2015} and spiral arms \citep{lyra2015}, as observed in several transition disks.  Under our assumption that the MHD wind and turbulence are independent, the disk mass and the accretion rates significantly decrease ($\sim0.001~M_{\odot}$ and $\lesssim 10^{-11}~M_{\odot}\rm{year}^{-1}$ at 5~Myr of evolution), and they become too low compared to observations of transition disks, as in the case of photo-evaporation \citep[e.g.][]{owen2011}. However, MHD winds are expected to carry away angular momentum and drive accretion \citep[e.g.][]{bai2013}

Different key observational diagnostics are needed to discern whether dead zones can be the origin of transition disk structures. First, one of the most promising methods is to spatially resolve the level of turbulence through the disk, measuring the non-thermal motions in the disk from spatially resolved molecular emission or measuring the broadening of molecular line profiles \citep[e.g.][]{carr2004, hughes2011, flaherty2015}. Second,  in the case of a dead zone, when a ring is  seen in the synthetic images of scattered optical light and in mm continuum emission, no clear radial difference exits for the cavity size between these wavelengths. Therefore, the combination of observations at optical or NIR scattered light and mm emission can help excluding the scenario where a dead zone alone acts on the disk. This is not the case when a dead zone and an MHD wind act together, however, which causes a clear segregation of small and large grains. Third,  observations of the gas distribution through CO and its isotopologues can highly contribute distinguishing a dead-zone scenario (with or without wind), in particular observations of the outer parts of the disk (farther out from the dust trap) where the dead-zone models predict a strong gas depletion. Future work will be devoted to a direct comparison of the results of this work with planet-disk interaction results, in particular predictions for CO, $^{13}$CO, and C$^{18}$O, to compare with upcoming observations with ALMA. In addition, future observations that can provide information about the distribution of the intermediate-size particles ($\sim100~\mu$m), with different instruments such as METIS, or ALMA through millimetre-polarisation \citep{pohl2016},  can bring significant insights for our interpretation of what causes the observational signatures of transition disks.

\begin{acknowledgements}
The authors are very thankful to C. Dominik and E.~F.~van Dishoeck for their valuable comments and lively discussions of this paper. We acknowledge X.-N~Bai, C.~Baruteau, S.~Facchini, G.~Rosotti, and C.~Walsh for their useful feedback, and we also thank the referees for their constructive reports. This work is supported by a Royal Netherlands Academy of Arts and Sciences (KNAW) professor prize. T.~B. acknowledges support from the DFG through SPP 1833 "Building a Habitable Earth" (KL 1469/13-1). 
\end{acknowledgements}

\appendix

\section{1D hydrodynamical simulations} \label{appendix_a}
We performed 1D hydrodynamical simulations including $\alpha$ as used in Eq.~\ref{dead_zone} using the PLUTO code \citep{mignone2012}. We used the same grid structure as was used for the previous results, with 300 cells in a logarithm grid, ranging from 1 to 200 AU. We used a third-order spatial and time-reconstruction method. The initial conditions follow model~\upperRomannumeral{2}. 

\begin{figure}
\centering
      	\includegraphics[width=9cm]{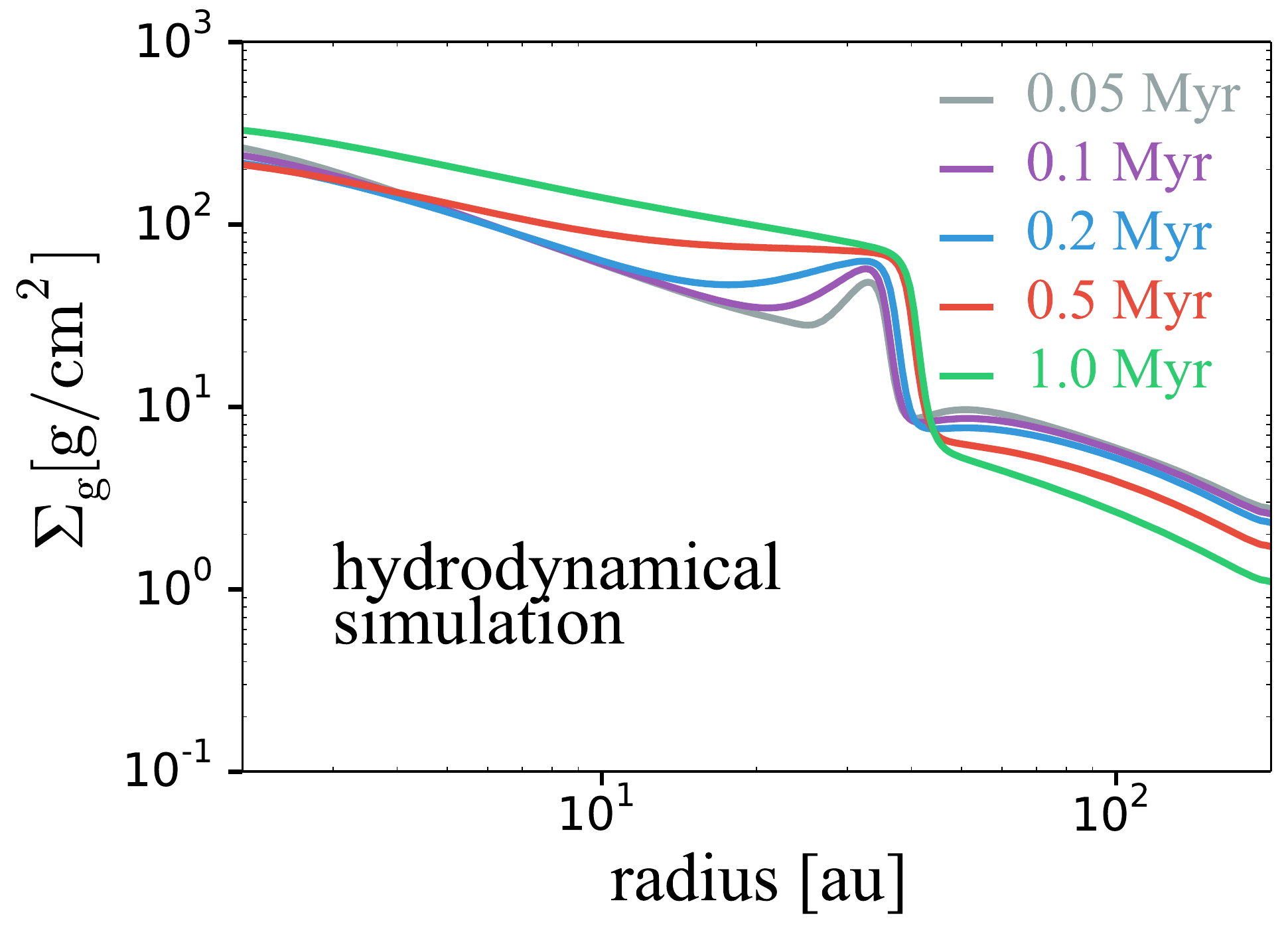}\\
	\includegraphics[width=9cm]{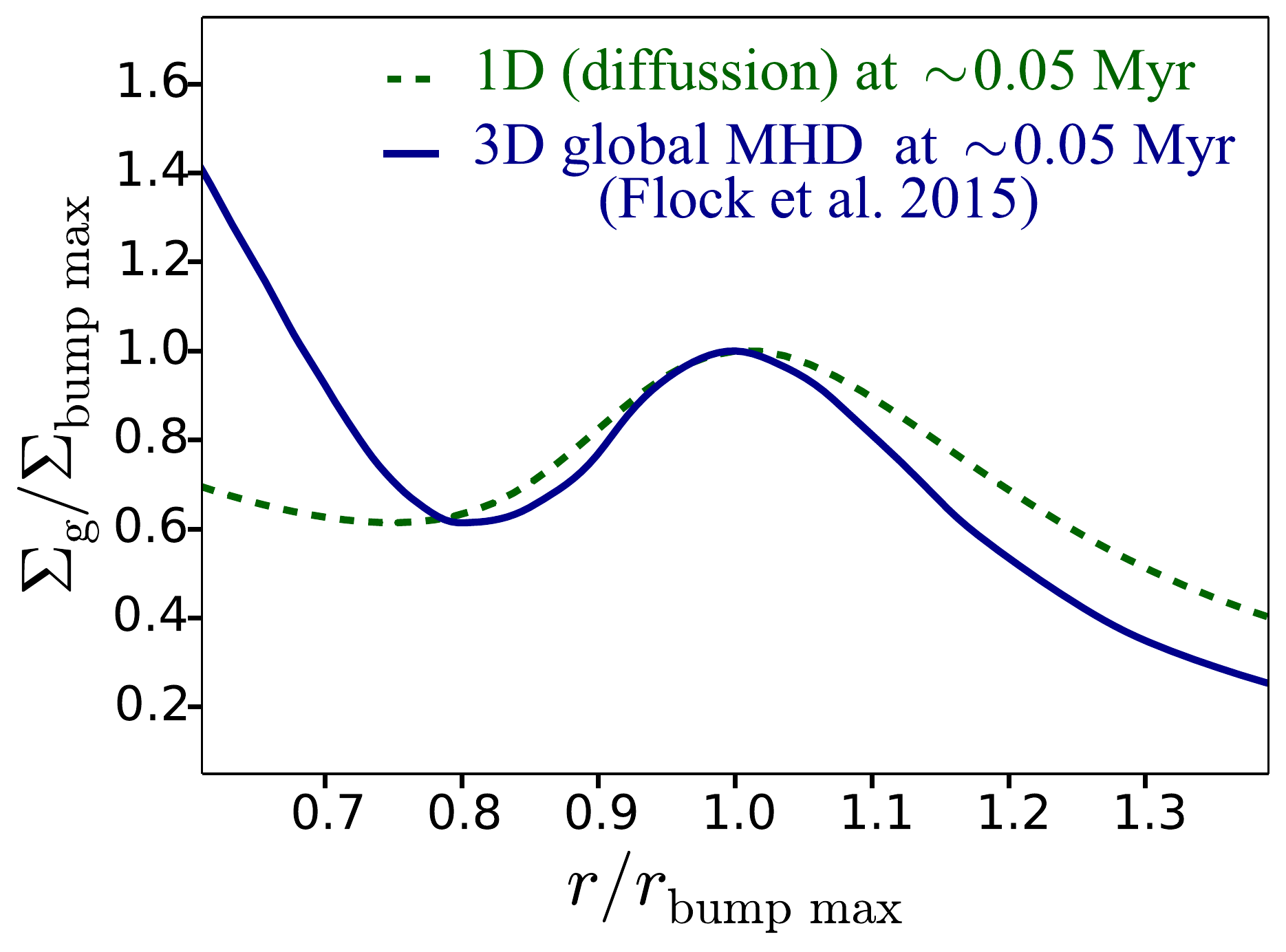}
\caption{Gas surface density evolution for different time snapshots for the hydrodynamical model~\upperRomannumeral{2} (top panel), and comparison of the 1D diffusion model with the 3D global MHD model from \cite{flock2015} at the same time of evolution $\sim0.05$~Myr (bottom panel), in normalised units with respect to the maximum of the bump formed at the edge of the dead zone.}
\label{fig:1DHD}
\end{figure}

The results are summarised in Fig.~\ref{fig:1DHD}. The results show that the  bump in the gas surface density vanishes on a shorter timescale in the hydrodynamical models than for the results in  Fig.~\ref{model1_2}. The differences  may arise because Eq.~\ref{eq_gas_evo} neglects that the gas moves with slightly  sub-Keplerian velocity, which can have an important effect when density gradients develop. However, we also compared the gas surface density profile from the 1D diffusion model with full 3D global stratified MHD models from \cite{flock2015}. At early times, the profiles look very similar in both models, showing a gap and jump structure. We note that in the 3D models, the bump in the gas surface density survives without significant changes in the amplitude for the whole simulation time ( $\gtrsim$0.1~Myr). It is possible that the inclusion of magnetic fields and the development of the RWI help to maintain the bump for longer times, and that the timescales shown in Fig.~\ref{model1_2} may not be underestimated too strongly when assuming more realistic simulations \citep[see also][]{regaly2012, miranda2016}.

\bibliographystyle{aa}

\end{document}